\documentclass[pre,preprint,showpacs,amsmath]{revtex4}

\usepackage{graphicx}
\usepackage{bm}% bold math

\renewcommand{\vec}[1]{\hm{#1}}
\newcommand{\cn}{\mbox{cn}}

\begin{document}

\title{Discrete breathers on symmetry-determined invariant manifolds for scalar models on the plane square lattice}
\author{G.S.Bezuglova}
\author{G.M. Chechin}
  \email{gchechin@gmail.com}
\author{P.P.Goncharov}
\affiliation{Department of Physics, Southern Federal University, Russia}
\date{\today}
\begin{abstract}

A group-theoretical approach for studying localized periodic and quasiperiodic vibrations in $2D$ and $3D$ lattice dynamical models is developed. This approach is demonstrated for the scalar models on the plane square lattice. The symmetry-determined invariant manifolds admitting existence of localized vibrations  are found and some types of discrete breathers are constructed on these manifolds.  A general method  using the apparatus of matrix representations of symmetry groups to simplify  the standard linear stability analysis is discussed. This method allows one to decompose the corresponding system of linear differential equations with time-dependent coefficients into a number of independent subsystems whose dimensions are less than the full dimension of the considered system.

\end{abstract}
\maketitle

\section*{Introduction}

The problem of energy localization in discrete {nonlinear} Hamiltonian systems attracts a considerable attention over the past few decades. In the framework of this problem a particular interest represent stationary {discrete breathers} (DBs) --- spatially localized and time-periodic dynamical objects. The history of discovery and discussion of different discrete breather  properties can be  found in a number of detailed review papers \cite{Aubry, Flach2, Aubry1, Flach3, Flach4}. Various analytical, numerical and experimental methods were  used to study breathers in a wide variety of discrete Hamiltonian systems. However, the most papers on discrete breathers deal with one-dimensional chains, and much smaller number of articles are devoted to study these dynamical objects in 2D and 3D periodic structures \cite{Fischer, Flach97,Butt-Wattis06,Butt-Wattis061,Feng,Kiselev-sievers,Kevrekidis,Doi,Xu-Qiang,Dmitriev,Dmitriev1,Yi-Wattis,Koukouloyannis}.

In the present paper, we discuss discrete breathers in 2D square lattice with one degree of freedom per lattice site. Such dynamical models are called {\it scalar}.

Different physical interpretation can be given to a scalar model. For example, it has been used for describing transversal mechanical vibrations of plane lattice in \cite{Fischer} (see Fig. 1), charge vibrations in an electrical network of nonlinear capacitors coupled to each other with linear inductors in \cite{Butt-Wattis061,Iran1} (see Fig. 2), etc.
\begin{figure}[htb]
\begin{center}
\includegraphics[scale=0.65]{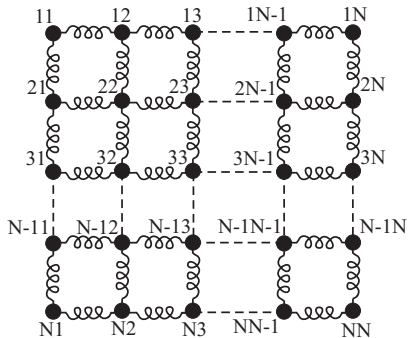}
\caption{Mechanical model.}{\label{fig01} }
\end{center}
\end{figure}

\begin{figure}[htb]
\begin{center}
\includegraphics[scale=0.65]{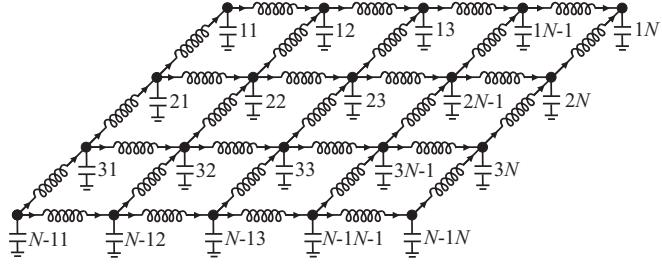}
\caption{Electrical model.}{\label{fig02} }
\end{center}
\end{figure}

In vector models, more than one  degree of freedom is associated with every lattice cite. For example, one can consider mechanical vibrations of the plane lattice with $x$- and $y$-displacements of the particles along its surface.

In many papers DBs in 2D and 3D lattices are investigated by some approximate methods such as "Rotating Wave Approximation"  (RWA) etc., and only several works are devoted to study these dynamical objects with the aid of numerically exact procedures. For example, in the paper \cite{Fischer}, stationary DBs were constructed for scalar models on square and hexagonal lattices with homogeneous interparticle potentials of different even degrees.

In the present paper, we develop a general group-theoretical method to obtain {\it invariant manifolds} admitting existence of localized excitations. Then the search of DBs can be done in two steps.

1. Singling out the above mentioned symmetry-determined invariant manifolds. To this end, we must know only the structure and the symmetry group of the considered physical system and we don't take into account  any information about interparticle interactions.

2. Constructing time-periodic dynamical regimes on each of these manifolds. Obviously such regimes represent stationary discrete breathers. Unlike the first step, we now need to know differential equations describing a given dynamical model.

Below we discuss the above approach using as examples the following two scalar dynamical models on the square lattice:

\begin{equation}\label{eq1}
      \begin{array}{l}
    \ddot{q}_{i,j}+\gamma\cdot q_{i,j}^{m-1}=(q_{i+1,j}-q_{i,j})^{m-1}-(q_{i,j}-q_{i-1,j})^{m-1}+\\
 \ \ \ \ \ \ \ \ \ \ \ \ \ \ \ \ \ \ \ \ \ \ \ \ \ \ +(q_{i,j+1}-q_{i,j})^{m-1}-(q_{i,j}-q_{i,j-1})^{m-1},
      \end{array}
\end{equation}

\begin{equation}\label{eq21}
      \begin{array}{l}
    \ddot{q}_{i,j}+\gamma\cdot q_{i,j}^3=q_{i+1,j}+q_{i-1,j}+q_{i,j+1}+q_{i,j-1}-4q_{i,j}.
      \end{array}
\end{equation}

For both models $i=1..M$, $j=1..N$ and periodic boundary conditions are assumed.  In equilibrium state all $q_{i,j}$ are equal to zero. Despite the fact that the dynamical variable $q_{i,j}$ associated with the site $(i,j)$  can be of different physical nature, we often refer to it as the particle (atomic) displacement.

Eqs.~(\ref{eq1}) describe dynamical model on square lattice whose "particles" interact with the nearest neighbours and with a substrate by homogeneous intersite and on-site potentials of the same degree $m$. Eqs.~(\ref{eq21}) determine the Klein-Gordon model, i.e. they describe an array of Duffing oscillators with linear interaction between the nearest neighbours. In both models, the strength of the on-site potential relative to the intersite potential is characterized by the coefficient $\gamma$, and we study {stability} of discrete breathers with respect to this parameter.

Note that the model (\ref{eq1}) was used in \cite{Fischer} without the on-site potential ($\gamma=0$).

Why we consider these two models? The model (\ref{eq1}) does not admit the harmonic approximation for $m>2$, and therefore it has no phonon spectrum. On the other hand, there exist exact solutions to this model representing Rosenberg nonlinear normal modes (NNMs) \cite{Rozenberg} (see, also, \cite{Mihlin-Manevich}). Existence of such solutions is {guaranteed} by the possibility to separate space and time variables as a consequence of {homogeneity} of the potential. In the dynamical regime corresponding to a given Rosenberg NNM, variables turn  out to be proportional to one and the same function of time $f(t)$, while spatial profile can be determined from a system of nonlinear \emph{algebraic} equations. Each  localized NNM represents a stationary discrete breather. Moreover, linear stability of these modes can be analyzed much simpler than the investigation based on the well-known Floquet method which one has to exploit in general case.

The second model possesses phonon spectrum and turns out to be more complicated for studying. Indeed, we must solve nonlinear \emph{differential} equations to construct discrete breathers and need in the Floquet method to analyze their stability.

It is essential that for both models we search breather solutions on the \emph{same symmetry-determined invariant manifolds} which are singled out by group-theoretical methods. The above invariant manifolds can be used to construct DBs for any other scalar models on the square lattice.

For vector models and/or for different lattices, one can find the corresponding symmetry-determined invariant manifolds using the same group-theoretical approach.

Let us emphasize that nonlinear differential equations of 2D and 3D dynamical models possess, as a rule, many different solutions and a symmetry-related classification is very desirable for any procedure of their construction.

Among solutions which can exist on the symmetry-determined invariant manifolds discussed in the present paper may be not only DBs, but various localized \emph{quasiperiodic} dynamical objects (some discussion of such objects in one-dimensional lattices can be found in \cite{ChechinDzh}).
The stability analysis of quasiperiodic regimes represents serious difficulties since one cannot use the Floquet method in such cases.

In \cite{Chechin-Zhukov}, we have developed a group-theoretical method for splitting the linearized dynamical equations near a given dynamical regime (periodic or quasiperiodic) with an arbitrary symmetry group. In some cases, this method allows one to simplify considerably the stability studying of a given dynamical regime, since high-dimensional linearized system is decomposed into a number of \emph{independent} subsystems of linear equations with time-dependent coefficients whose dimensions can be sufficiently small.

The present paper possesses the following structure. The group-theoretical method for finding symmetry-determined invariant manifolds is discussed in Sec.~\ref{Sec1}.  Discrete breathers in scalar model (\ref{eq1}) with homogeneous potentials of $m=4, \ 6, \ 8$ degrees are considered in Sec.~\ref{Sec2}, while their stability is analyzed in Sec.~\ref{Sec3}. Construction and stability of DBs in the Klein-Gordon model 2 are discussed in Sec.~\ref{Sec4}. The group-theoretical method for simplifying stability analysis of periodic and quasiperiodic regimes in dynamical systems with discrete symmetry is presented in Sec.~\ref{Sec5}. The results of the paper are briefly summarized in Conclusion.

\section{Symmetry-determined invariant manifolds}{\label{Sec1}}

\subsection{General discussion}

All possible solutions to a given system of differential equations $L$ can be classified by \emph{subgroups} $G_j$ of the "parent" symmetry group $G_0$ consisting of all transformations of dynamical variables which transform this system to its equivalent form. In other worlds, $G_0$ is the group of \emph{invariance} of the considered dynamical model. In particular case of a Hamiltonian system, $G_0$ can be treated as a group of all transformations under which its Hamiltonian turns out to be invariant.

If we take a certain solution to the dynamical equations $L$ and begin to act on it by all transformations $g\in G_0$, then some of these transformations conserve the solution --- they form a subgroup $G\in G_0$. On the other hand, acting  on the above solution by
other $g\in G_0$, i.e. by $g\not\in G$, we obtain the \emph{orbit} of $m$ exact solutions to which our original solution belongs. The integer number $m$ is the \emph{index} of the subgroup $G_j$ in the group $G_0$.

Thus, for the symmetry classification of all possible dynamical regimes in the given model with parent group $G_0$, we must consider all its subgroups $G_j\in G_0$. Each $G_j$ determines an \emph{invariant manifold} in the phase space of the dynamical system: if we select initial conditions belonging to this manifold and begin to integrate the differential equations of our model, we obtain a dynamical regime which never leaves the above manifold during the time evolution. Invariant manifolds corresponding to subgroups of the symmetry group of a given model we call "symmetry-determined invariant manifolds".

The idea of classification of  dynamical regimes by symmetry-determined invariant manifolds was taken as a principle in the theory of \emph{bushes of nonlinear normal modes} in physical systems with discrete symmetry \cite{DAN1, PhysD98} (see also \cite{Columbus}). A given $m$-dimensional bush is described by $m$ dynamical variables and represents an \emph{exact} solution belonging to a certain symmetry-determined invariant manifold of the considered dynamical model.

In \cite{DAN1, PhysD98, FPU1, FPU2, PhysD, Chechin-Zhukov}, we have obtained low-dimensional bushes of {extended} vibrational modes for dynamical models on various space-periodic structures. For example, in \cite{FPU1} this has been done for all possible 3D crystal structures described by any of 230 space groups. We have studied existence and stability of one-dimensional and two-dimensional  bushes of NNMs in the FPU chains in \cite{FPU2, PhysD}.

The group-theoretical approach for finding bushes of NNMs and for analyzing their stability properties was described in detail in \cite{Columbus}. However, up to the present time, we applied this approach to study bush existence and stability for the case of \emph{extended} NNMs. For this purpose, we need to make use of the subgroups $G_j$ of the parent space group $G_0$ which include some \emph{translational} symmetry elements. The primitive cell of the resulting spatial pattern of the {vibrational} state is some times greater than the primitive cell of the equilibrium state.

In the present paper, we consider stationary discrete breathers which represent \emph{localized} vibrational regimes. As a consequence, they are associated with \emph{point} subgroups of the corresponding parent space group.

\subsection{Choice of the parent symmetry group}

To begin the procedure of finding symmetry-determined invariant manifolds, we must choose a parent symmetry group $G_0$ which embraces all transformations of dynamical variables under which the considered mathematical model is invariant. Taking into account that the stationary state represents one of the solutions to the model dynamical equations and that, therefore, the symmetry of this state must be a certain subgroup of $G_0$, we, firstly, consider the spatial transformations under which the square lattice is invariant.

It is well known, that the symmetry of square lattice is characterized by the plane space group $C_{4v}^1=P4mm$ (in Schoenflies and in the International notations, respectively) and we can assume $G_0=C_{4v}^1$.

In general, the parent group $G_0$ can be wider than the symmetry group of the stationary state. For example, several stationary states with the same energy and symmetry can correspond to a given Hamiltonian system and, as a consequence, the symmetry elements which transform these stationary states into each other must also enter the group $G_0$.

Moreover, the parent group $G_0$ can contain some additional symmetry elements due to the specific type of interparticle interactions in the considered model. For both models (\ref{eq1}), (\ref{eq21}), all forces acting on particles turn out to be \emph{odd} functions of the atomic displacements $q_{ij}$ (the corresponding potentials represent even functions of their arguments). As a consequence, changing signs of all dynamical variables, which we denote by symbol $\hat{P}$,  also represents a transformation under which models (\ref{eq1}), (\ref{eq21}) are invariant and $\hat{P}$ must be incorporated into the parent group $G_0$.

Therefore, in our case, the symmetry parent group $G_0$ can be written as the direct  product of the space group $C_{4v}^1$ and the group($E$, $\hat{P}$):
\begin{equation}\label{eq10}
    G_0=C_{4v}^1\otimes (E, \ \hat{P}).
\end{equation}
The last group consists of two elements, $E$ and $\hat{P}$, where $E$ is identity element.

For scalar models on square lattice which are not invariant with respect to the transformation $\hat{P}$, we can assume $G_0=C_{4v}^1$.

If one find any additional transformations which don't change a given mathematical model, they can also be incorporated into the parent symmetry group $G_0$. It such a case, we will obtain more detailed symmetry-related classification of the dynamical regimes because some more subgroups correspond to the wider parent group.

\subsection{Wyckoff positions}{\label{labRPS}}

As was already discussed, for finding invariant manifolds associated with stationary discrete breathers, we must deal with \emph{point} subgroups of the parent symmetry group $G_0$ (\ref{eq10}). Such subgroups (they don't contain any translational symmetry elements), can be found for any of 230 space groups in the standard textbooks on crystallography. One can find there a list of so called Wyckoff positions (WPs) or {regular point sets} in the primitive crystal cell which correspond to point subgroups of a given space group. WPs determine points in the crystalline lattice with different \emph{local} symmetry and, in any crystal, atoms are distributed among these points.

However, WPs characterize the local symmetry of  different points in crystal lattice in equilibrium state. On the other hand, for the vibrational regime corresponding to a given DB, we deal with atoms displaced from their equilibrium positions. As a consequence,  for a fixed instant, we have an atomic pattern whose symmetry is characterized by a subgroup of the local symmetry group of the crystal equilibrium state. Thus, we must find {all subgroups} of the crystal local (point) symmetry groups.

For the square lattice with plane group $C_{4v}^1$, there are six WPs (see, e.g., \cite{Kovalev}) and only three of them must be considered for our purpose. In the International notation, they read:
\begin{equation}\label{eq11a}
    1a, \ 4mm \ (0,0);
\end{equation}

\begin{equation}\label{eq11b}
    1b, \ 4mm \ \Bigl(\frac{1}{2}, \frac{1}{2}\Bigr);
\end{equation}

\begin{equation}\label{eq11c}
2c, \ mm \  \Bigl(0, \frac{1}{2}\Bigr).
\end{equation}
Hereafter, we will refer to these Wyckoff positions as WP-1, WP-2, and WP-3, respectively. The following information is given for each WP in (\ref{eq11a})-(\ref{eq11c}): the multiplicity $(1; \ 1; \ 2)$ and the label $(a; \ b; \ c)$, the local point group $(4mm; \ 4mm; \ mm)$ and the localization of the fixed point of this group in square primitive cell with unit edge \ $\Bigl[(0,0); \ \Bigl(\frac{1}{2}, \frac{1}{2}\Bigr);  \Bigl(0, \frac{1}{2}\Bigr)\Bigr]$.

The local groups of other Wyckoff positions turn out to be subgroups of the local groups listed in (\ref{eq11a})-(\ref{eq11c}), and we will automatically take them into account when select {all} subgroups of the groups specified in (\ref{eq11a})-(\ref{eq11c}).

\begin{figure}[htb]
\begin{center}
\includegraphics[scale=1]{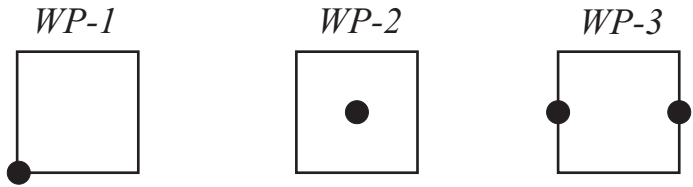}
\caption{Wyckoff positions in the primitive cell of square lattice.
}{\label{figRPS} }
\end{center}
\end{figure}

The WP-1 and WP-2 possess identical point symmetry groups ($4mm$), but these groups are \emph{located} in \emph{different ways} with respect to the square cell of the considered lattice: the fixed  point of the  group $4mm$ of WP-1 lies at the corner of the primitive cell, while the group $4mm$ of WP-2 located in the center of the cell (see, Fig.~\ref{figRPS}). Because of this reason, subgroups of corresponding point groups of WP-1 and WP-2 generate \emph{different} invariant manifolds (see Table 1).

The point group ($mm$) of WP-3 is located at the middle of the cell edge.

Note that WP-1 and WP-2 consist of only one point per primitive cell (their multiplicity is unity), while WP-3 consists of {two} points per the cell (see, Fig.~\ref{figRPS}).

\subsection{Construction of symmetry-determined invariant manifolds}

In general case, dynamical equations of a given scalar model on the square lattice can be written in the form
\begin{equation}\label{eqv1}
    \ddot{\vec{Q}}_{M\times N}=\vec{F}(\vec{Q}_{M\times N}).
\end{equation}
Here matrix
\begin{equation}\label{eqv2}
    \vec{Q}_{M\times N}=\left(\begin{array}{cccc}
                    q_{11} & q_{12} & ... & q_{1N} \\
                    q_{21} & q_{22} & ... & q_{2N} \\
                    ... & ... & ... & ... \\
                    q_{M1} & q_{M2} & ... & q_{MN} \\
                  \end{array}\right)
\end{equation}
determines the set of all variables $q_{ij}=q_{ij}(t)$ corresponding to the $M\times N$ fragment of the
lattice, while matrix $\vec{F}$ represents r.h.s. of the differential equations of the considered
model (examples of such models are given by Eqs.~(\ref{eq1}) and (\ref{eq21})).

As was discussed in Sec.~\ref{labRPS}, we must consider all subgroups of local symmetry groups of WP-1,
WP-2 and WP-3 demanding that the matrix (pattern) $\vec{Q}_{M\times N}$ is invariant with respect to these subgroups. As a result, we will obtain all possible symmetry-determined invariant manifolds.

The restriction on $\vec{Q}_{M\times N}$ to be invariant under the action of a symmetry group $G\subset G_0$ leads to certain relations between the dynamical variables $q_{ij}(t)$: some of them turn out to be equal or differ from each other only in sign. Below, presenting a symmetry determined invariant manifold, we denote equal dynamical variables by the same symbols $a(t)$, \ $b(t)$, \ $c(t)$, \ $d(t)$ ... and don't point out explicitly their time-dependence.

In the present paper, we study only \emph{strongly localized} discrete breathers. This allows us to deal with relatively small fragments $\vec{Q}_{M\times N}$ of each invariant manifold. We choose concrete values of $M$ and $N$ from the condition that amplitudes of $q_{ij}(t)$ for {peripheral} sites must be much smaller than those for the breather core.

Let us demonstrate a procedure for finding $3\times 3$ manifold $Q^{(1)}_{3\times 3}$ invariant with respect to the point group $G=C_{4v}$.

This group consists of eight elements: four of them ($g_1$, \ $g_2$, \ $g_3$, \ $g_4$) are rotations by the angles $0^{\circ}$, \ $90^\circ$, \ $180^\circ$, \ $270^\circ$ around $z$ axis  orthogonal to the lattice plane, while other elements ($g_5$, \ $g_6$, \ $g_7$, \ $g_8$) are reflections in the mirror planes orthogonal to 3D vectors (1,0,0), (1,1,0), (0,1,0), (-1,1,0). This four reflections we denote by the symbols $\sigma_x$, $\sigma_y$, $\sigma_{xy}$, $\sigma_{-xy}$, respectively. [Detailed description of the group $C_{4v}$ one can find below in Eqs.~(\ref{eqBB100}).]

Above listed symmetry elements, acting on the manifold
\begin{equation}\label{eq15}
    \vec{Q}^{(1)}_{3\times 3}=\left(\begin{array}{ccc}
                    q_{11} & q_{12} & q_{13} \\
                    q_{21} & q_{22} & q_{23} \\
                    q_{31} & q_{32} & q_{33} \\
                  \end{array}\right),
\end{equation}
transpose its elements $q_{ij}\equiv q_{ij}(t)$ in a certain way. In particular, for symmetry element $g_2$, representing rotation by $90^\circ$ around $Z$ axis passing through the center of the square cell, we have
\begin{equation}\label{eq16}
    \hat{g}_2 \vec{Q}^{(1)}_{3\times 3}=\left(\begin{array}{ccc}
                    q_{13} & q_{23} & q_{33} \\
                    q_{12} & q_{22} & q_{32} \\
                    q_{11} & q_{21} & q_{31} \\
                  \end{array}\right),
\end{equation}
while for $g_5$, representing reflection $\sigma_x$, we obtain
\begin{equation}\label{eq17}
    \hat{g}_5 \vec{Q}^{(1)}_{3\times 3}=\left(\begin{array}{ccc}
                    q_{13} & q_{12} & q_{11} \\
                    q_{23} & q_{22} & q_{21} \\
                    q_{33} & q_{32} & q_{31} \\
                  \end{array}\right).
\end{equation}

Note that $g_2$ and $g_5$ can be considered as \emph{generators} of the group $G=C_{4v}$: all other elements of this group can be expressed via different products of $g_2$ and $g_5$.

As a consequence, to obtain manifold, invariant with respect to the group $G=C_{4v}$, we can require that this manifold is invariant relative to $g_2$ and $g_5$ only:
\begin{equation}\label{eq20}
    \hat{g}_2 \vec{Q}^{(1)}_{3\times 3}=\vec{Q}^{(1)}_{3\times 3},
\end{equation}
\begin{equation}\label{eq211}
    \hat{g}_5 \vec{Q}^{(1)}_{3\times 3}=\vec{Q}^{(1)}_{3\times 3}.
\end{equation}
Then from Eqs.~(\ref{eq20}), (\ref{eq15}) and (\ref{eq16}), we obtain
\begin{eqnarray*}
% \nonumber to remove numbering (before each equation)
     q_{12}=q_{21}=q_{32}=q_{23}=b(t), \\
     q_{11}=q_{31}=q_{33}=q_{13}=c(t), \\
     q_{22}= a(t).
\end{eqnarray*}
Here we have introduced three new variables $a(t)$, $b(t)$ and $c(t)$, instead of nine old variables \ $q_{ij}(t)$, \ $i=1..3, \ j=1..3$.

Similarly, we must take into account invariance condition (\ref{eq211}) comparing Eqs.~(\ref{eq15}) and (\ref{eq17}). As a result, we obtain the following form of the manifold $\vec{Q}^{(1)}_{3\times 3}$ invariant with respect to the point group $G=C_{4v}$:
\begin{equation}\label{eq50a}
    \vec{Q}^{(1)}_{3\times 3}=\left(\begin{array}{ccc}
                    c & b & c \\
                    b & a & b \\
                    c & b & c \\
                  \end{array}\right)
\end{equation}

Let us note that Eq.~(\ref{eq211}) \emph{does not} produce any additional restrictions on the dynamical variables as compared to Eq.~(\ref{eq20}), i.e. invariant manifold (\ref{eq50a}) is actually determined only by Eq.~(\ref{eq20}). This means that above obtained manifold $\vec{Q}^{(1)}_{3\times 3}$ turns out to be invariant not only relative to the group \ $C_4=\{C_4\}$ with one generator $g_2$, representing rotation by $90^\circ$, but simultaneously relative to its supergroup $C_{4v}=\{C_4, \sigma_x\}$ with two generators --- $g_2$ and $g_5$. (Hereafter, we define any symmetry group by the list of its generators given in curle braces).

However, if we consider \emph{larger} fragment of dynamical variables on square lattice, for example, the manifold $\vec{Q}^{(1)}_{5\times 5}$, Eq.~(\ref{eq211}) indeed produce additional restrictions on some variables $q_{ij}$ and, as a consequence, the manifolds invariant with respect to the groups
$C_4=\{C_4\}$ and $C_{4v}=\{C_4, \sigma_x\}$ prove to be different (see, $\vec{Q}^{(1)}$ and $\vec{Q}^{(2)}$ in Table 1).

Among point subgroups of the parent symmetry group, there are such that generate invariant manifolds admitting existence of localized, as well as delocalized dynamical objects. Indeed, the manifold (\ref{eq50a}) allows localized periodic or quasiperiodic modes if $|a|>>|b|>>|c|$, while the manifold
\begin{equation*}
    \left(\begin{array}{ccc}
                    b & a & c \\
                    b & a & c \\
                    b & a & c \\
                  \end{array}\right)
\end{equation*}
does not admit localization in Y-direction because variables $q_{ij}$, expressed via $a$, $b$, $c$, have no tendency to decrease by amplitude from its center to the periphery.

Note that some invariant manifolds can correspond to multibreathers of complex structure, for example, with zero amplitude in the breather center. In Table 1, we present all invariant manifolds associated with subgroups of the parent symmetry group (\ref{eq10}) which admit construction of {\it simple} discrete breathers --- they are localized in both $X$- and $Y$- directions and have no zero in the center.

However, at the end of each WP section of Table 1, we give the list of all other subgroups associated with this regular point set. Such information can be used for construction some different types of invariant manifolds, corresponding to this WP.

Moreover, below each manifold from Table~1, we give its symmetry group which can be useful to \emph{expand} a given manifold in the case of bad discrete breather localization (vibrational amplitudes slightly decrease from the breather core to its periphery).

In Table 1, we present only small fragments of invariant manifolds $\vec{Q}^{(j)}$ \ ($j=1..27$), for example, for manifolds corresponding to WP-1 we show $5\times 5$ fragments and give their $3\times 3$ parts in frames.

Note that manifolds   with a minus sign before some variables correspond to subgroups whose elements contain operator $\hat{P}$ (independently or in combination with other symmetry elements). We list these manifolds after those corresponding to the parent group $G_0=C_{4v}$. The former can be used only for models  with even potential, while the latter correspond to the models with an arbitrary potential.

Finally, let us emphasize once more that all invariant manifolds presented in Table 1 admit existence of periodic, as well as \emph{quasiperiodic} dynamical objects.

In the following sections of the paper, we discuss construction of DBs as spatially localized and time-periodic vibrations on some manifolds listed in Table 1 for the dynamical models (\ref{eq1}) and (\ref{eq21}).

\newpage

\hfill{Table 1}

{Symmetry-determined invariant manifolds for scalar models on the square lattice.
}

{\centerline{\textbf{WP-1}}}

{\small{

\noindent
\begin{tabular}{|c|c|c|c|c|}
  \hline
  % after \\: \hline or \cline{col1-col2} \cline{col3-col4} ...
 $  \left(
      \begin{array}{ccccc}
        f & e & d & e & f \\
            \cline{2-4}
        e & \multicolumn{1}{|c}{c} & b & \multicolumn{1}{c|}{c} & e \\
        d & \multicolumn{1}{|c}{b} & a & \multicolumn{1}{c|}{b} & d \\
        e & \multicolumn{1}{|c}{c} & b & \multicolumn{1}{c|}{c} & e \\
            \cline{2-4}
        f & e & d & e & f \\
      \end{array}
    \right)$
&
$ \left(
      \begin{array}{ccccc}
        g & e & d & f & g \\
            \cline{2-4}
        f & \multicolumn{1}{|c}{c} & b & \multicolumn{1}{c|}{c} & e \\
        d & \multicolumn{1}{|c}{b} & a & \multicolumn{1}{c|}{b} & d \\
        e & \multicolumn{1}{|c}{c} & b & \multicolumn{1}{c|}{c} & f \\
            \cline{2-4}
        g & f & d & e & g \\
      \end{array}
    \right)  $
&
 $ \left(
      \begin{array}{ccccc}
        h & f & e & g & j \\
            \cline{2-4}
        f & \multicolumn{1}{|c}{c} & b & \multicolumn{1}{c|}{d} & g \\
        e & \multicolumn{1}{|c}{b} & a & \multicolumn{1}{c|}{b} & e \\
        g & \multicolumn{1}{|c}{d} & b & \multicolumn{1}{c|}{c} & f \\
            \cline{2-4}
        j & g & e & f & h \\
      \end{array}
    \right)$
&
 $ \left(
      \begin{array}{ccccc}
        m & k & f & e & n \\
            \cline{2-4}
        h & \multicolumn{1}{|c}{d} & b & \multicolumn{1}{c|}{e} & j \\
        g & \multicolumn{1}{|c}{c} & a & \multicolumn{1}{c|}{c} & g \\
        j & \multicolumn{1}{|c}{e} & b & \multicolumn{1}{c|}{d} & n \\
            \cline{2-4}
        n & e & f & k & m \\
      \end{array}
    \right)$
&
  $ \left(
      \begin{array}{ccccc}
        j & h & e & h & j \\
            \cline{2-4}
        g & \multicolumn{1}{|c}{d} & b & \multicolumn{1}{c|}{d} & g \\
        f & \multicolumn{1}{|c}{c} & a & \multicolumn{1}{c|}{c} & f \\
        g & \multicolumn{1}{|c}{d} & b & \multicolumn{1}{c|}{d} & g \\
            \cline{2-4}
        j & h & e & h & j \\
      \end{array}
    \right)  $
    \\

  $Q^{(1)}\!: \ \{C_4, \sigma_y\}$  & $Q^{(2)}\!: \ \{C_4$\} & $Q^{(3)}\!: \ \{C_2, \sigma_{xy}$\}& $Q^{(4)}\!: \ \{C_2$\} &$Q^{(5)}\!: \ \{C_2, \sigma_y$\} \\
  \hline

 $  \left(
      \begin{array}{ccccc}
        n & i & g & j & p \\
            \cline{2-4}
        i & \multicolumn{1}{|c}{d} & b & \multicolumn{1}{c|}{e} & k \\
        g & \multicolumn{1}{|c}{b} & a & \multicolumn{1}{c|}{c} & h \\
       j & \multicolumn{1}{|c}{e} & c & \multicolumn{1}{c|}{f} & l \\
            \cline{2-4}
        p & k & h & l & m \\
      \end{array}
    \right)$
&
 $  \left(
      \begin{array}{ccccc}
        n & j & i & k & p \\
            \cline{2-4}
        g & \multicolumn{1}{|c}{e} & b & \multicolumn{1}{c|}{f} & m \\
        j & \multicolumn{1}{|c}{c} & a & \multicolumn{1}{c|}{d} & h \\
        g & \multicolumn{1}{|c}{e} & b & \multicolumn{1}{c|}{f} & m \\
            \cline{2-4}
        n & j & i & k & p \\
      \end{array}
    \right)$
&
 \multicolumn{3}{p{9cm}|}{

$\{C_4 \hat{P}\}$, \ \ $\{C_2 \hat{P}\}$, \ \ $\{\sigma_y \hat{P}\}$, \ \ $\{\sigma_{xy} \hat{P}\}$, \ \  $\{C_4 \hat{P}, \sigma_y\}$,

$\{C_4 \hat{P}, \sigma_y \hat{P}\}$, \ \ $\{C_2, \sigma_y \hat{P}\}$, \ \  $\{C_2, \sigma_{xy} \hat{P}\}$,

 $\{C_2 \hat{P}, \sigma_y\}$, \ \  $\{C_2 \hat{P}, \sigma_{xy}\}$, \ \ $\{C_4, \sigma_y \hat{P}\}$}

\\

  $Q^{(6)}\!: \ \{\sigma_{xy}$\}  & $Q^{(7)}\!: \ \{\sigma_{y}$\} &  \multicolumn{3}{c|}{ } \\
\hline

\end{tabular}}}

\bigskip

{\small{

{\centerline{\textbf{WP-2}}}

\hspace{-1cm}
\begin{tabular}{|c|c|c|c|c|}
  \hline
  % after \\: \hline or \cline{col1-col2} \cline{col3-col4} ...
  $ \left(
      \begin{array}{cccc}
        a & b & b & a \\
\cline{2-3}
        b & \multicolumn{1}{|c}{c} & \multicolumn{1}{c|}{c} & b \\
        b & \multicolumn{1}{|c}{c} & \multicolumn{1}{c|}{c} & b \\
\cline{2-3}
        a & b & b & a \\
      \end{array}
    \right)  $
&
  $ \left(
      \begin{array}{cccc}
        a & b & b & a \\
 \cline{2-3}
        c & \multicolumn{1}{|c}{d} & \multicolumn{1}{c|}{d} & c \\
        c & \multicolumn{1}{|c}{d} & \multicolumn{1}{c|}{d} & c \\
 \cline{2-3}
        a & b & b & a \\
      \end{array}
    \right)  $
&
  $ \left(
      \begin{array}{cccc}
        a & b & c & a \\
\cline{2-3}
        c & \multicolumn{1}{|c}{d} & \multicolumn{1}{c|}{d} & b \\
        b & \multicolumn{1}{|c}{d} & \multicolumn{1}{c|}{d} & c \\
\cline{2-3}
        a & c & b & a \\
      \end{array}
    \right)  $
&
  $ \left(
      \begin{array}{cccc}
        a & b & c & d \\
      \cline{2-3}
        b & \multicolumn{1}{|c}{e} & \multicolumn{1}{c|}{f} & c \\
        c & \multicolumn{1}{|c}{f} & \multicolumn{1}{c|}{e} & b \\
      \cline{2-3}
        d & c & b & a \\
      \end{array}
    \right)  $
&
  $ \left(
      \begin{array}{cccc}
        a & b & c & d \\
      \cline{2-3}
        e & \multicolumn{1}{|c}{f} & \multicolumn{1}{c|}{g} & h \\
        h & \multicolumn{1}{|c}{g} & \multicolumn{1}{c|}{f} & e \\
      \cline{2-3}
        d & c & b & a \\
      \end{array}
    \right)  $

    \\

  $Q^{(8)}\!: \ \{C_4, \sigma_y\}$  & $Q^{(9)}\!: \ \{C_2, \sigma_y$\} & $Q^{(10)}\!: \ \{C_4$\} & $Q^{(11)}\!: \ \{C_2, \sigma_{xy}$\} &$Q^{(12)}\!: \ \{C_2$\}

  \\

\hline

  $ \left(
      \begin{array}{cccc}
        a & b & c & d \\
      \cline{2-3}
        b & \multicolumn{1}{|c}{e} & \multicolumn{1}{c|}{f} & g \\
        c & \multicolumn{1}{|c}{f} & \multicolumn{1}{c|}{k} & l \\
      \cline{2-3}
        d & g & l & p \\
      \end{array}
    \right)  $

    &

      $ \left(
      \begin{array}{cccc}
        a & b & c & d \\
            \cline{2-3}
        e & \multicolumn{1}{|c}{f} & \multicolumn{1}{c|}{g} & h \\
        e & \multicolumn{1}{|c}{f} & \multicolumn{1}{c|}{g} & h \\
            \cline{2-3}
        a & b & c & d \\
      \end{array}
    \right)  $

&

  $ \left(
      \begin{array}{cccc}
        a & b & -b & -a \\
      \cline{2-3}
        c & \multicolumn{1}{|c}{d} & \multicolumn{1}{c|}{-d} & -c \\
        c & \multicolumn{1}{|c}{d} & \multicolumn{1}{c|}{-d} & -c \\
      \cline{2-3}
        a & b & -b & -a \\
      \end{array}
    \right)  $

&
   $ \left(
      \begin{array}{cccc}
        a & b & c & -a \\
      \cline{2-3}
        -c & \multicolumn{1}{|c}{d} & \multicolumn{1}{c|}{-d} & -b \\
        -b & \multicolumn{1}{|c}{-d}& \multicolumn{1}{c|}{d} & -c \\
      \cline{2-3}
        -a & c & b & a \\
      \end{array}
    \right)  $
&

  $ \left(
      \begin{array}{cccc}
        a & b & -b & -a \\
  \cline{2-3}
        b & \multicolumn{1}{|c}{c} & \multicolumn{1}{c|}{-c} & -b \\
        -b & \multicolumn{1}{|c}{-c} & \multicolumn{1}{c|}{c} & b \\
\cline{2-3}
        -a & -b & b & a \\
      \end{array}
    \right)  $

\\

   $Q^{(13)}\!: \ \{\sigma_{xy}$\}  & $Q^{(14)}\!: \ \{\sigma_{y}$\} &   $Q^{(15)}\!: \  \{C_2\hat{P}, \sigma_y$\}  &  $Q^{(16)}\!: \ \{C_4 \hat{P}$\} &  $Q^{(17)}\!: \ \{C_4 \hat{P}, \sigma_{y} \hat{P}$\}  \\
\hline

  $ \left(
      \begin{array}{cccc}
        a & b & -b & -a \\
      \cline{2-3}
        c & \multicolumn{1}{|c}{d} & \multicolumn{1}{c|}{-d} & -c \\
        -c & \multicolumn{1}{|c}{-d} & \multicolumn{1}{c|}{d} & c \\
      \cline{2-3}
        -a & -b & b & a \\
      \end{array}
    \right)  $

    &

  $ \left(
      \begin{array}{cccc}
        a & b & c & d \\
            \cline{2-3}
        e & \multicolumn{1}{|c}{f} & \multicolumn{1}{c|}{g} & h \\
        -e & \multicolumn{1}{|c}{-f} & \multicolumn{1}{c|}{-g} & -h  \\
            \cline{2-3}
        -a & -b & -c & -d \\
      \end{array}
    \right)  $
&

  $ \left(
      \begin{array}{cccc}
        a & b & c & d \\
            \cline{2-3}
        e & \multicolumn{1}{|c}{f} & \multicolumn{1}{c|}{g} & h \\
        -h & \multicolumn{1}{|c}{-g} & \multicolumn{1}{c|}{-f} & -e  \\
            \cline{2-3}
        -d & -c & -b & -a \\
      \end{array}
    \right)  $

&

\multicolumn{2}{p{7cm}|}{

$\{\sigma_{xy} \hat{P}\}$, \ \ $\{C_2, \sigma_{xy} \hat{P}\}$, \ \ $\{C_2 \hat{P}, \sigma_{xy}\}$,

 $\{C_4, \sigma_y\}$, \ \ $\{C_4, \sigma_y \hat{P}\}$.}

\\

  $Q^{(18)}\!: \ \{C_2, \sigma_y \hat{P}$\}  & $Q^{(19)}\!: \ \{\sigma_{y} \hat{P}$\} &   $Q^{(20)}\!: \ \{C_2 \hat{P}$\}  &    \multicolumn{2}{c|}{ } \\
\hline

\end{tabular}}}

\bigskip

\newpage

{\centerline{\textbf{WP-3}}}

\noindent
\begin{tabular}{|c|c|c|c|}
  \hline
  % after \\: \hline or \cline{col1-col2} \cline{col3-col4} ...
  $ \left(
      \begin{array}{ccccc}
        a & b & c & d & e \\
            \cline{2-4}
        f & \multicolumn{1}{|c}{g} & h & \multicolumn{1}{c|}{j}  & k \\
        f & \multicolumn{1}{|c}{g} & h & \multicolumn{1}{c|}{j}  & k \\
            \cline{2-4}
        a & b & c & d & e \\
      \end{array}
    \right)  $
&
 $  \left(
      \begin{array}{ccccc}
        a & b & c & b & a \\
            \cline{2-4}
        f & \multicolumn{1}{|c}{g} & h & \multicolumn{1}{c|}{g}  & f \\
        f & \multicolumn{1}{|c}{g} & h & \multicolumn{1}{c|}{g}  & f \\
            \cline{2-4}
        a & b & c & b & a \\
      \end{array}
    \right)$
&
 $ \left(
      \begin{array}{ccccc}
        a & b & c & b & a \\
            \cline{2-4}
        f & \multicolumn{1}{|c}{g} & h & \multicolumn{1}{c|}{g}  & f \\
        l & \multicolumn{1}{|c}{m} & n & \multicolumn{1}{c|}{m}  & l \\
            \cline{2-4}
        p & q & r & q & p \\
      \end{array}
    \right) $
&
 $ \left(
      \begin{array}{ccccc}
        a & b & c & d & e \\
            \cline{2-4}
        f & \multicolumn{1}{|c}{g} & h & \multicolumn{1}{c|}{j}  & k \\
        k & \multicolumn{1}{|c}{j} & h & \multicolumn{1}{c|}{g} & f \\
            \cline{2-4}
        e & d & c & b & a \\
      \end{array}
    \right)$

\\

  $Q^{(21)}\!: \ \{\sigma_y\}$  & $Q^{(22)}\!: \ \{C_2, \sigma_y\}$ & $Q^{(23)}\!: \ \{\sigma_x\}$ & $Q^{(24)}\!: \ \{C_2\}$  \\

  \hline

 $  \left(
      \begin{array}{ccccc}
        a & b & c & d & e \\
            \cline{2-4}
        f & \multicolumn{1}{|c}{g} & h &  \multicolumn{1}{c|}{j} & k \\
        -f & \multicolumn{1}{|c}{-g} & -h &  \multicolumn{1}{c|}{-j} & -k \\
            \cline{2-4}
        -a & -b & -c & -d & -e \\
      \end{array}
    \right)$
&
$  \left(
      \begin{array}{ccccc}
        a & b & c & b & a \\
            \cline{2-4}
        f & \multicolumn{1}{|c}{g} & h & \multicolumn{1}{c|}{g} & f \\
        -f & \multicolumn{1}{|c}{-g} & -h & \multicolumn{1}{c|}{-g} & -f \\
            \cline{2-4}
        -a & -b & -c & -b & -a \\
      \end{array}
    \right)$
&
 $  \left(
      \begin{array}{ccccc}
        a & b & c & d & e \\
            \cline{2-4}
        f & \multicolumn{1}{|c}{g} & h &  \multicolumn{1}{c|}{j} & k \\
        -k & \multicolumn{1}{|c}{-j} & -h &  \multicolumn{1}{c|}{-g} & -f \\
            \cline{2-4}
        -e & -d & -c & -b & -a  \\
      \end{array}
    \right)$
&

\multicolumn{1}{p{4cm}|}{

$\{\sigma_x \hat{P}\}$, \ \ $\{\sigma_y,  \sigma_x \hat{P}\}$,

$\{\sigma_y \hat{P}, \sigma_x \hat{P}\}$

}

\\

  $Q^{(25)}\!: \ \{\sigma_y \hat{P}\}$  & $Q^{(26)}\!: \ \{\sigma_y \hat{P}, \sigma_{x}\}$ & $Q^{(27)}\!: \ \{C_2 \hat{P}\}$   &
\multicolumn{1}{c|}{ }  \\
\hline

\end{tabular}

\section{Construction of discrete breathers for the model with homogeneous potential}{\label{Sec2}}

Each invariant manifold from Table 1 depends on a number of arbitrary parameters ($a$, $b$, $c$, ...). To construct a discrete breather, we must find such values of these parameters which lead to a spatially localized and {time-periodic vibration} when they are used as initial values for integrating dynamical equations of a given mathematical model.

In this section, we discuss construction of DBs for the model~(\ref{eq1}) which corresponds to the case of homogeneous potential of $m$-degree. Similar model \emph{without on-site} potential and without studying breather stability was considered in \cite{Fischer}. Unlike this paper, we study the dependence of existence and stability of DBs with respect to the relative strength ($\gamma$) of on-site and intersite parts of the potential energy of the model~(\ref{eq1}).

The specific structure of Eqs.~(\ref{eq1}) admits the space-time separation and, as a consequence, we can treat discrete breathers for the case of the homogeneous potential of $m$ degree in terms of localized nonlinear normal modes (NNMs) introduced by Rosenberg \cite{Rozenberg}. To this end, we assume that \begin{equation}\label{eq30b}
q_{i,j}(t)=k_{ij}\cdot f(t),
\end{equation}
where \ $i=1..M$, \ $j=1..N$, while $k_{ij}$ are constant coefficients.

Substituting the ansatz (\ref{eq30b}) into differential equations (\ref{eq1}) and requiring that all they are equivalent to each other, we obtain  a number of nonlinear \emph{algebraic} equations, which determine the spatial profile of DB, and one ("governing") differential equation, which determines time-dependence of all the dynamical variables $q_{ij}(t)$.  This time-dependence is described by the single function $f(t)$.

If we now take into account particular structures of the invariant manifolds from Table~1, the number of unknown coefficients $k_{ij}$ will be equal to the number of manifold arbitrary parameters $a$, $b$, $c$, ... minus one, since one of these parameters can be fixed.

In \cite{ChechinDzh, ChechinBez}, we have used Rosenberg modes technique for constructing DBs and studying their stability in one-dimensional lattices. In the present paper, we use the same technique to investigate DBs in the scalar model (\ref{eq1}) on the two-dimensional square lattice.

Let us illustrate the procedure of the Rosenberg mode construction using as an example $\vec{Q}^{(1)}_{3\times 3}$ invariant manifold (\ref{eq50a}). Only three independent dynamical variables, $a(t)$, $b(t)$ and $c(t)$, correspond to this manifold. According to the definition of Rosenberg modes, we assume that these variables are proportional to the same time-dependent function $f(t)$:

$a(t)=a\cdot f(t)$, \ \ \ $b(t)=b\cdot f(t)$, \ \ \ $c(t)=c\cdot f(t)$.

\noindent Here $a$, $b$, $c$ in r.h.s. of the equalities are {constants} which we use, for simplicity, instead of the coefficients $k_{ij}$ from Eq.~(\ref{eq30b}). Therefore, the Rosenberg mode $R(t)$ can be written in the form:

\begin{equation}\label{eq35b}
    R(t)=\left(\begin{array}{ccc}
                    c & b & c \\
                    b & a & b \\
                    c & b & c \\
                  \end{array}\right)\cdot f(t),
\end{equation}
where the vibrational amplitude corresponding to the center site of the manifold can be assumed equal  to unity ($a=1$). Substituting dynamical variables $q_{ij}(t)$ from (\ref{eq35b}) into differential equations (\ref{eq1}), we can obtain the following algebraic equations

\begin{equation}\label{eq31a}
     b\cdot[-\gamma+4(b-1)^{m-1}]=-\gamma \cdot b^{m-1}+2\cdot(c-b)^{m-1}+(1-b)^{m-1},\\
\end{equation}

\begin{equation}\label{eq31b}
     c\cdot[-\gamma+4(b-1)^{m-1}]=-\gamma \cdot c^{m-1}+2\cdot(b-c)^{m-1},
\end{equation}
while the governing equation reads
\begin{equation}\label{eq32b}
    \ddot{f}(t)+p^2\cdot f(t)^{m-1}=0, \ \ \ \ p^2=\gamma-4\cdot(b-a)^{m-1}.
\end{equation}

As a rule, there are many different solutions to Eqs.~(\ref{eq31a}), (\ref{eq31b}) and we must demand that $|a|>|b|>|c|$ to provide a localized breather profile.

There are many methods for solving nonlinear algebraic equations, such as various versions of the Newton-Rafson method, the steepest descent method, etc \cite{Flach4}. We use for this purpose the standard procedure of the mathematical package MAPLE. In this way, we have obtained breathers on a number of invariant manifolds presented in Table 1 for different degrees $m$  of the homogeneous potential and for a dense set of values $\gamma\in[0; \ 10]$.

Let us consider some examples.

1. For the homogeneous potential of $m=4$ degree, we have obtained the following breather spatial profiles on the invariant manifold $\vec{Q}^{(1)}_{3\times 3}$ for different values of $\gamma$:
\begin{equation}\label{eq100b}
    \begin{array}{c}
                    \gamma=0: \ \ \ \ a=1, \ \ b=-0.25439, \ \ c=0.00439;  \\
                    \gamma=3: \ \ \ \ a=1, \ \ b=-0.17354, \ \ c=0.00113; \\
                    \gamma=6: \ \ \ \ a=1, \ \ b=-0.12261, \ \ c=0.00032. \\
                  \end{array}
\end{equation}

We see that the coefficients $b$ and $c$ decrease with increasing $\gamma$, i.e. breather localization becomes better with increasing the strength of the on-site potential with respect to the intersite potential. (Note that the breather for $\gamma=0$ coincides with that found in \cite{Fischer}).

The time-dependence of the breather solutions with spatial profiles (\ref{eq100b}) determined by the governing equation
\begin{equation}\label{eqB2b}
    \ddot{f}(t)+p^2\cdot f(t)^3=0, \ \ \ \ p^2=4(a-b)^3+\gamma,
\end{equation}
with analytical solution of the form
\begin{equation}\label{eqB21b}
    f(t)=\cn\Bigl(\omega t, \frac{1}{\sqrt{2}}\Bigr), \ \ \ \ \omega=p\cdot f(0).
\end{equation}

\begin{figure}[htb]
\begin{center}
\includegraphics[scale=0.8]{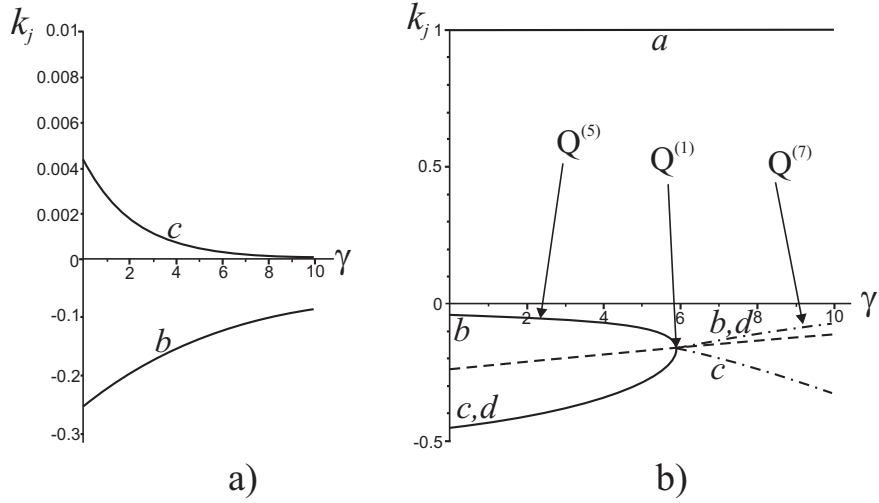}
\caption{Discrete breathers for the model 1 on the invariant manifolds a) $Q^{(1)}_{3\times 3}$; \ b) $Q^{(1)}_{3\times 3}$,  $Q^{(5)}_{3\times 3}$,  $Q^{(7)}_{3\times 3}$.}{\label{fig4} }
\end{center}
\end{figure}

In Fig.~\ref{fig4}a, we show the coefficients $b=b(\gamma)$ and $c=c(\gamma)$ as functions of the parameter $\gamma\in[0; 10]$. Note that these functions possess \emph{opposite signs} and are plotted on different scales.

 Thus, we can speak about the \emph{breather family} because there exists a certain DB for any fixed value of the parameter $\gamma$. Moreover, one can see from Fig.~\ref{fig4}a that $b(\gamma)$ and $c(\gamma)$ represent, in the considered case, sufficiently simple functions which can be approximated by some polynomials with good accuracy. Using the mean-square method we find the following approximations:
 \begin{equation}\label{eqB10}
    \begin{array}{c}
                    b(\gamma)=-0.133940\cdot10^{-2}\gamma^2+0.297771\cdot10^{-1}\gamma-0.252179,\\
                    c(\gamma)=-0.980247\cdot10^{-5}\gamma^3+0.213750\cdot10^{-3}\gamma^2-0.158541\cdot10^{-2}\gamma+0.423062\cdot10^{-2}.\\
                  \end{array}
\end{equation}

The above discussed breather family on the manifold $Q^{(1)}_{3\times 3}$ for $m=4$, \ $\gamma\in [0; \ 10]$ is described by two variables \ $b(\gamma)$, \ $c(\gamma)$ because the third variable, $a$ (see Table 1), can be considered as a constant: \ $a=1$. Another fixed value of this variable leads to the breather spacial profile which turns out to be proportional to that for $a=1$ (despite the algebraic equations (\ref{eq31a}), (\ref{eq31b})  for finding Rosenberg nonlinear normal mode profiles are nonlinear!).

However,  the time dependence of the breather solution, in particular, its frequency, does depend on the parameter $a$, since $p=p(a)$ according to Eq.~(\ref{eqB2b}).

Let us represent profiles for some other breathers whose stability we will discuss in the next section.

a) Invariant manifold \ $Q^{(1)}_{3\times 3}$ \ $\{C_4, \sigma_y\}$.

 \begin{equation}\label{eqB11}
    \begin{array}{c}
                    m=6: \ a(\gamma)=1,\\
                    \phantom{m=6:} \  b(\gamma)=-0.733310\cdot10^{-3}\gamma^2+0.228270\cdot10^{-1}\gamma-0.251758,\\
                    \phantom{m=6:} \  c(\gamma)=-0.346820\cdot10^{-6}\gamma^3+0.780072\cdot10^{-5}\gamma^2-0.593555\cdot10^{-4}\gamma+0.157500\cdot10^{-3}.\\
                  \end{array}
\end{equation}

\begin{equation}\label{eqB12}
    \begin{array}{c}
                    m=8: \ a(\gamma)=1,\\
                    \phantom{m=8:} \  b(\gamma)=-0.132486\cdot10^{-3}\gamma^2+0.151641\cdot10^{-1}\gamma-0.251854,\\
                    \phantom{m=8:} \  c(\gamma)=-0.103097\cdot10^{-7}\gamma^3+0.255021\cdot10^{-6}\gamma^2-0.215981\cdot10^{-5}\gamma+0.636736\cdot10^{-5}.\\\\
                  \end{array}
\end{equation}

b) Invariant manifold \ $Q^{(8)}_{3\times 3}$ \ $\{C_4, \sigma_y\}$.

\begin{equation}\label{eqB13}
    \begin{array}{c}
                    m=4: \ a(\gamma)=1,\\
                    \phantom{m=4:} \  b(\gamma)=0.206049\cdot10^{-4}\gamma^3-0.462088\cdot10^{-2}\gamma^2+0.864977\cdot10^{-1}\gamma-0.539657,\\
                    \phantom{m=4:} \  c(\gamma)=-0.155129\cdot10^{-3}\gamma^3+0.332604\cdot10^{-2}\gamma^2-0.232426\cdot10^{-1}\gamma+0.538939\cdot10^{-1}\\
                  \end{array}
\end{equation}

\begin{equation}\label{eqB14}
    \begin{array}{c}
                    m=6: \ a(\gamma)=1,\\
                    \phantom{m=6:} \  b(\gamma)=-0.334370\cdot10^{-3}\gamma^3+0.420770\cdot10^{-2}\gamma^2+0.272984\cdot10^{-1}\gamma-0.500903\\
                    \phantom{m=6:} \  c(\gamma)=-0.541969\cdot10^{-5}\gamma^3+0.153732\cdot10^{-3}\gamma^2-0.143670\cdot10^{-2}\gamma+0.445843\cdot10^{-2}\\
                  \end{array}
\end{equation}

\begin{equation}\label{eqB15}
    \begin{array}{c}
                    m=8: \ a(\gamma)=1,\\
                    \phantom{m=8:} \  b(\gamma)=0.104004\cdot10^{-3}\gamma^3+0.174863\cdot10^{-3}\gamma^2+0.151320\cdot10^{-1}\gamma-0.500456,\\
                    \phantom{m=8:} \  c(\gamma)=0.161985\cdot10^{-6}\gamma^3+0.133814\cdot10^{-5}\gamma^2-0.751460\cdot10^{-4}\gamma+0.460920\cdot10^{-3}\\
                  \end{array}
\end{equation}

b) Invariant manifold \ $Q^{(15)}_{3\times 3}$ \ $\{C_2 \hat{P}, \sigma_y\}$.

\begin{equation}\label{eqB16}
    \begin{array}{c}
                    m=4: \ a(\gamma)=1,\\
                    \phantom{m=4:} \  b(\gamma)=0.479180\cdot10^{-4}\gamma^3-0.131456\cdot10^{-2}\gamma^2+0.161263\cdot10^{-1}\gamma-0.1356280688,\\
                    \phantom{m=4:} \  c(\gamma)=-0.971222\cdot10^{-6}\gamma^3+0.208977\cdot10^{-4}\gamma^2-0.155620\cdot10^{-3}\gamma+0.442289\cdot10^{-3}.\\
                  \end{array}
\end{equation}

\begin{equation}\label{eqB17}
    \begin{array}{c}
                    m=6: \ a(\gamma)=1,\\
                    \phantom{m=6:} \  b(\gamma)=-0.259370\cdot10^{-4}\gamma^2+0.113473\cdot10^{-2}\gamma-0.344276\cdot10^{-1},\\
                    \phantom{m=6:} \  %c(\gamma)=0.1649903663e-11\gamma^3-0.4625131291e-10\gamma^2+0.5302559625e-9\gamma-0.2813907816e-8.
c(\gamma)\sim 10^{-7}.
\\
                  \end{array}
\end{equation}

\begin{equation}\label{eqB18}
    \begin{array}{c}
                    m=8: \ a(\gamma)=1,\\
                    \phantom{m=8:} \  b(\gamma)=-0.479130\cdot10^{-6}\gamma^2+0.659596\cdot10^{-4}\gamma-0.813363\cdot10^{-2}, \\
                    \phantom{m=8:} \  %c(\gamma)=-0.5806223935e-19\gamma^2+0.2241526737e-17\gamma-0.3612654693e-16.
c(\gamma)\sim 10^{-16}.
\\
                  \end{array}
\end{equation}

 In Fig.~\ref{fig4}b, one can see an example of nontrivial profiles of DBs for the homogeneous potential of $m=8$ degree. These breathers are constructed on the invariant manifold $Q^{(7)}_{3\times 3}$ which is characterized by five arbitrary parameters $b$, $c$, $d$, $e$, $f$ (the parameter $a=1$ corresponds to the breather centre). Parameters $e$ and $f$ corresponding to the lattice sites more distant from the central site with $a=1$ are very  small ($e$, $f\sim10^{-6}$) {as compared} to parameters $b$, $c$, $d$ associated with the nearest sites.

 There is a critical point \ $\gamma=\gamma_c\approx5.885$ in which some DBs of \emph{different symmetry} coincide with each other. Indeed, the dotted line on which $b=c=d$ corresponds to  the breathers on the invariant manifolds $Q^{(1)}_{3\times 3}$ with the symmetry group $\{C_4, \ \sigma_y\}\equiv C_{4v}$. The solid lines ($c=d\neq b$) corresponds to breathers on the manifold $Q^{(5)}_{3\times 3}$ with the symmetry group $\{C_2, \ \sigma_y\}\equiv C_{2v}$, while the dash-dot lines ($b=d\neq c$) corresponds to DBs on the manifold $Q^{(7)}_{3\times 3}$ whose symmetry is described by the group $\{\sigma_y\}\equiv C_s$.

 Note that the above mentioned breathers, except for those who possess the symmetry group $\{C_4, \ \sigma_y\}\equiv C_{4v}$, turn out to be unstable (see below).

\section{Stability analysis of discrete breathers for the model with homogeneous potential}{\label{Sec3}}

Discrete breathers represent time-periodic dynamical regimes and, therefore, standard Floquet method can be used for analyzing their stability. However, for the scalar model (\ref{eq1}) on the square lattice with $m$-degree homogeneous potential the stability studying of discrete breathers can be essentially simplified, if they are {Rosenberg nonlinear normal modes} (in the next section we will consider breathers in the model 1 which are not such modes!). In \cite{ChechinBez, ChechinDzh}, we have demonstrated this idea for the case of one-dimensional chains with $m=4$ homogeneous potential, while now we use the same method for analyzing the breather stability in the two-dimensional scalar model (\ref{eq1}) described by the homogeneous potential of arbitrary (even) degree $m$.

Let us consider a \emph{localized} nonlinear normal mode. Being periodic in time, this dynamical object represents a certain discrete breather. Let vector $\vec{\delta}(t)$ determines a set of infinitesimal deviations of all the particles from the exact breather solution $R(t)$ [for example, see Eq.~(\ref{eq100b})]. Linearizing nonlinear differential equations of the model (\ref{eq1}) with respect to all components of the vector $\vec{\delta}$, we obtain  the following system of linear differential equations with time-periodic coefficients (some computational details can be found in \cite{ChechinDzh}]):
\begin{equation}\label{eqB200}
    \ddot{\vec{\delta}}=(m-1)\cdot f^{m-2}\cdot \vec{B}\cdot \vec{\delta}.
\end{equation}
Here $\vec{B}$ is a constant symmetric matrix whose dimension, as well as the dimension of the vector $\vec{\delta}$, is equal to the number of all sites of a chosen lattice fragment,  while $f(t)$ is the solution to the governing equation for the Rosenberg NNM.

For the model (\ref{eq1}), described by the homogeneous potential of $m$ degree, the time-periodic function $f(t)$ is the solution to the equation
\begin{equation}\label{eqBB10}
    \ddot{f}(t)+p^2\cdot f(t)^{m-1}=0,
\end{equation}
with initial conditions
\begin{equation}\label{eqBB12}
    f(0)=A, \ \ \ \ \dot{f}(0)=0.
\end{equation}
Here, $A$ is the breather amplitude (displacement of the central particle), while \ $p^2$ depends on the breather profile. For example, for a breather on the invariant manifold $Q^{(1)}_{3\times 3}$ in the case of the model 1 with $m=4$, we have Eq.~(\ref{eqB2b}).

In the above approach, investigation of stability of the considered dynamical regime (discrete breather, in our case) is reduced to analyzing the stability of  the \emph{zero solution} of the linearized variational system (\ref{eqB200}).

Note that the system (\ref{eqB200}) possesses a very specific structure: it contains the constant matrix $B$ with time-periodic coefficient $(m-1)\cdot f^{m-2}(t)$ standing in front of it.

Now we can reduce the symmetric matrix $B$ to a diagonal form $D$ with the aid of a certain orthogonal transformation
\begin{equation}\label{eqB205}
    \vec{S}^T \vec{B} \vec{S} = \vec{D}.
\end{equation}
($\vec{S}^T$ is a transpose of $\vec{S}$).

Eigenvalues $\lambda_j$ of the matrix $\vec{B}$ are diagonal elements of the matrix $D$, while eigenvectors of $\vec{B}$ form the columns of the matrix $\vec{S}$.

Introducing a new infinitesimal vector $\vec{z}$ instead of the old one $\vec{\delta}$ with the aid of the transformation $\vec{\delta}=\vec{S}\vec{z}$,
we decompose the system (\ref{eqB200}) of coupled equations into \emph{independent} scalar equations of one and the same type:
\begin{equation}\label{eqB206}
    \ddot{z}=(m-1)\cdot \lambda_j\cdot f(t)^{m-2} \cdot z.
\end{equation}

With the help of scaling
\begin{equation}\label{eqB207}
f(t)=A\psi(t), \ \ \ t=\frac{\tau}{p A^{m/2-1}}
\end{equation}
one can rewrite Eqs.~(\ref{eqBB10}), (\ref{eqBB12}) and (\ref{eqB206}) in the form
\begin{equation}\label{eqB208}
\psi''_{\tau}+\psi^{m-1}(\tau)=0,
\end{equation}
\begin{equation}\label{eqB208a}
\psi(0)=1, \ \ \ \ \psi'_{\tau}(0)=0,
\end{equation}
\begin{equation}\label{eqB209}
z''_{\tau}=\psi^{m-2}(\tau)\cdot \Lambda_j \cdot z,
\end{equation}
\begin{equation}\label{eqB209a}
z(0)=0, \ \ \ \ z'_{\tau}(0)=0,
\end{equation}
where
\begin{equation}\label{eqB210}
\Lambda_j=\frac{(m-1)\cdot \lambda_j}{p^2}
\end{equation}

Here we denote differentiation by the new time variable $\tau$  by prime. Eqs.~(\ref{eqB209a}) appear because we study stability of the zero solution of the differential equation (\ref{eqB209}) for  the function $z(t)$.

We will refer to the parameters $\Lambda_j$ as \emph{stability indicators}. Breather solution proves to be stable if all $\Lambda_j$ fall into intervals of stability of the zero solution to Eq.~(\ref{eqB209}).

Note that one of the stability indicators, say $\Lambda_0$, always belongs to the boundary between the first region of instability and the second region of stability of zero solution to Eq.~(\ref{eqB209}). The infinitesimal vector corresponding to this indicator turns out to be exactly proportional to the breather profile (i.e., it is located along this profile) and, therefore, such indicator $\Lambda_0$ does not affect the breather stability.

For $m=4$, stability analysis of zero solution of Eq.~(\ref{eqB209}) can be fulfilled with the aid of the Lame equation (see, \cite{ChechinDzh}). For $m>4$, such analysis represents essentially more complicated analytical problem \cite{Mihlin}. However, some numerical methods can be efficiently used for studying stability of the zero solution of Eq.~(\ref{eqB209}). Using these methods we were able to reveal that discrete breathers prove to be linear stable if all indicators $\Lambda_j$ fall into the following intervals:
\begin{equation}\label{eqB211}
[0; \ 1], \ \ [m-1; \ m+2], \ \ [3m-2; \ 3m+3], ...
\end{equation}

\begin{figure}[htb]
\begin{center}
\includegraphics[scale=0.99]{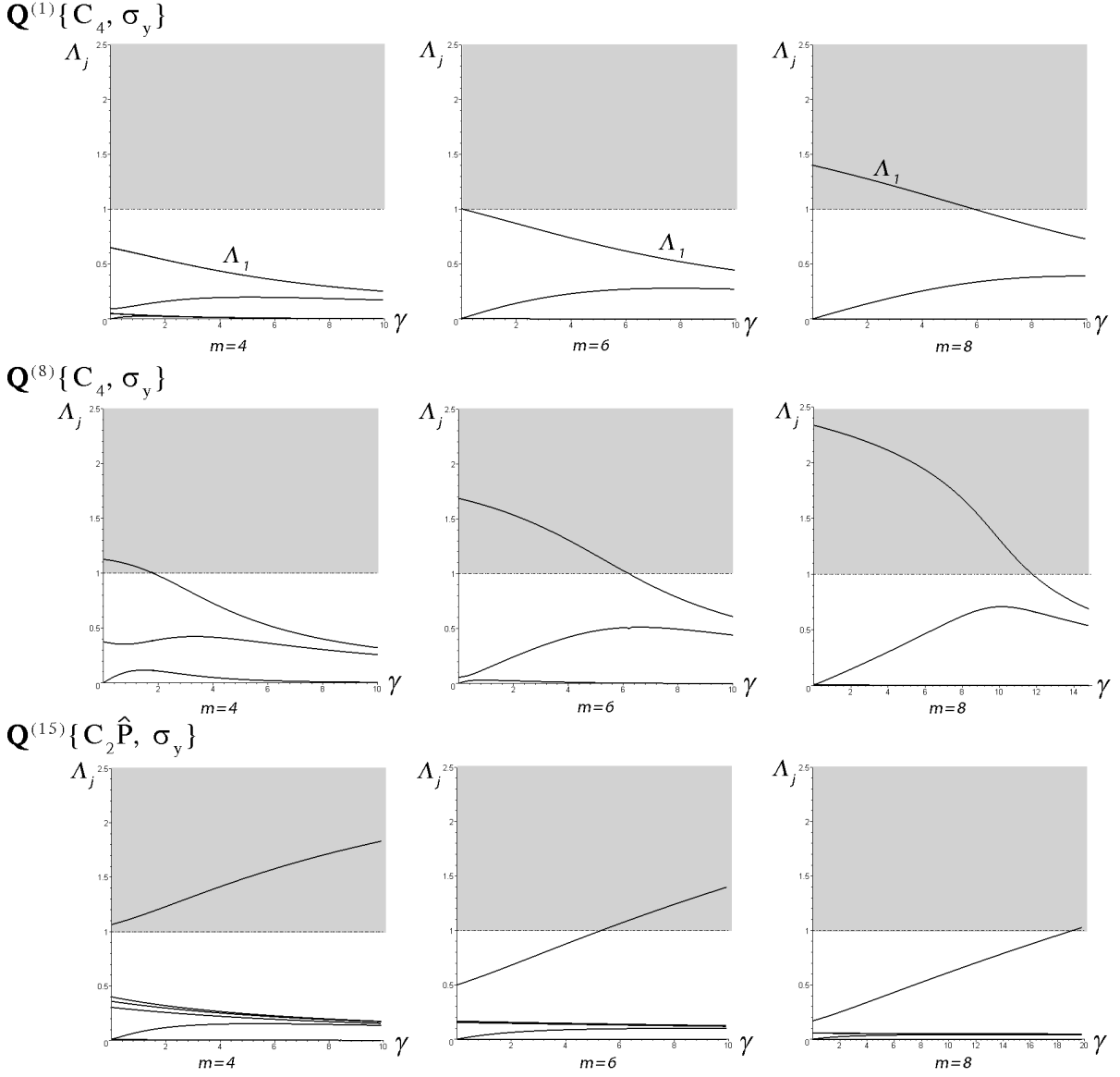}
\caption{Stability indicators $\Lambda_j(\gamma)$ for discrete breathers on the invariant manifolds $\vec{Q}^{(1)} \ \{C_4, \sigma_y\}$, \ $\vec{Q}^{(8)} \ \{C_4, \sigma_y\}$, \ $\vec{Q}^{(15)} \ \{C_2 \hat{P}, \sigma_y\}$ for the model 1.}{\label{fig5} }
\end{center}
\end{figure}

Let us consider stability of discrete breathers listed in Eqs.~(\ref{eqB10})--(\ref{eqB18}). The indicators $\Lambda_j=\Lambda_j(\gamma)$ are certain functions of the parameter $\gamma$ describing relative strength of the on-site and intersite potentials of the model 1. In Fig.~\ref{fig5}, we present functions $\Lambda_j(\gamma)$, \ $\gamma\in[0; \ 10]$ for different manifolds $\vec{Q}^{(i)}$ and different degrees $m$ of the homogeneous potential.

Note that stable and unstable regions of the zero solution of Eq.~(\ref{eqB209}) \emph{alternate} according to Eq.~(\ref{eqB211}). The first stable region $[0, \ 1]$  are one and the same for all $m$, while the width of the first unstable region $[1, m-1]$ increases with increasing $m$. Because of this reason, we depict in Fig.~\ref{fig5} only lower part  of the first region of instability (it is presented in grey color).

Let us  comment in these figures.

1) DBs corresponding to the invariant manifold $Q^{(1)}_{3\times3}$ with the symmetry group $\{C_4, \sigma_y\} = C_4v$ centered at the corner of the primitive cell (WP-1) are described for $m = 4, 6, 8$ by Eqs. (\ref{eqB10}--\ref{eqB12}). Nine degree of freedom are associated with this $3\times3$ fragment of the square lattice and, therefore, there are nine indicators $\Lambda_j$  which determine stability of breathers on $Q^{(1)}_{3\times3}$.

For example, for $m = 4$, $\gamma = 0$, we have found the following indicators $\Lambda_j$ ($j = 0, 1,..., 8$):
 \begin{equation}\label{eq250}
    \begin{array}{c}
                    \lambda_0=3, \ \  \lambda_1=\lambda_2=0.65095, \ \ \lambda_3=0.64879\\
                    \lambda_4=0.09128, \ \ \lambda_5=0.05089, \ \  \lambda_6=\lambda_7=0.04873, \ \ \lambda_8\sim10^{-10}\\
                  \end{array}
\end{equation}

Note that some indicators $\Lambda_j$ are equal to each other (we consider the source of such degeneration elsewhere). The functions $\Lambda_j = \Lambda_j(\gamma)$ are depicted for the breathers on $Q^{(1)}_{3\times3}$ for $m = 4, 6, 8$ \ $\gamma\in[0; 10]$ in Fig~\ref{fig5}. For $m = 4$, all stability indicators $\Lambda_j(\gamma)$ fall into the first stability region since $0 < \Lambda_j(\gamma) < 1$, $j = 0 .. 8$, $\gamma\in[0; 10]$. Note that we cannot see all these indicators in Fig.~\ref{fig5} because some of them are too small. Therefore, discrete breathers described by Eq.~(\ref{eqB10}) ($m = 4$) are stable dynamical objects for all values of the parameter $\gamma$ on the considered interval.

However, one can see essentially different stability properties of discrete breathers for $m = 8$ on $Q^{(1)}_{3\times 3}$. Indeed, one curve, $\Lambda_{1}(\gamma)$, lies out of the first stability region for $\gamma < \gamma_c \approx 5.8848$. This means  that DBs for small strength of the on-site potential turn out to be unstable. On the other hand, for $\gamma > \gamma_c$, this curve enters the first stability region and DBs become stable for $\gamma$ greater than $\gamma_c$.

Finally, for $m = 6$, we have an intermediate case between two above discussed cases: the curve $\Lambda_{1}(\gamma)$ begins exactly from the upper end of the first stability region and, therefore, the breather possesses marginal stability for $\gamma = 0$. All DBs for $\gamma > 0$ are stable since $\Lambda_j(\gamma)<1$, $j = 0 .. 8$, $\gamma\in(0;10]$.

2) Another situation takes place for DBs (\ref{eqB13} -- \ref{eqB15}) on the invariant manifold $Q^{(8)}_{3\times 3}$ with the group $\{C_4; \sigma_y\} = C_4v$ corresponding to the center of the primitive cell (WP-2). Indeed, we see in Fig.~\ref{fig5} that for $m = 4$, \ $m = 6$ and $m = 8$, some stability indicators lie outside the first stability region up to certain critical values $\gamma_c$ of the parameter $\gamma$. These critical values $\gamma_c = \gamma_c(m)$ are different for different $m$. Moreover, we see that $\gamma_c(m)$ increases with increasing $m$. In other words, we need more strong on-site potential for stabilizing DB for the greater $m$.

3) Finally, for DBs determined by Eqs. (\ref{eqB16})--(\ref{eqB18}) on the invariant manifold $Q^{(15)}_{4\times4}$ with the symmetry group $\{C_2 \hat{P}; \sigma_y\} \equiv C_2v$ corresponding to WP-2, we find $\Lambda_j(\gamma)$, $j = 0 .. 15$ which also are depicted in Fig.~\ref{fig5}. From these plots, one can see that DBs for $m = 4$ are unstable for all values of $\gamma\in[0; 10]$. On the other hand, for $m = 6$ and $m = 8$, DBs turn out to be stable for small values of $\gamma$ up to the critical values $\gamma_c^{(1)} \approx 5.3104$ and $\gamma_c^{(2)} \approx 19.1248$, respectively. In all above discussed cases, there is the same tendency: a sufficiently strong on-site potential can stabilize breather vibrations.

In conclusion, let us pay attention to one important property of the discrete breather stability for the models with homogeneous potential of an arbitrary degree $m$. In such models, the stability of the breathes which represent Rosenberg nonlinear normal modes \emph{don't depend} on their amplitude. This fact becomes obvious, if we take into account that the amplitude
$A$ does {not enter} Eqs.~(\ref{eqB208})--(\ref{eqB210}) which fully determine the stability of the zero solution to decoupled linear equations derived from the system (\ref{eqB200}).

\section{Discrete breathers for arbitrary scalar models on the square lattice}{\label{Sec4}}

\subsection{Construction of DBs for the model 2}

The scalar model 1 with homogeneous potential of $m > 2$ degree discussed in the previous section represents a sufficiently exotic case from the physical point of view. Indeed, there is no phonon spectrum in this model since it does not admit the harmonic approximation: Taylor-series expansions of r.h.s. of Eqs. (\ref{eq1}) with respect to all dynamical variables $q_{ij}$ don't contain any linear terms. On the other hand, in the more realistic models, admitting the harmonic approximation, there are no {\it localized} Rosenberg nonlinear normal modes. Therefore, we have to use different and more complicated methods, as compared to those described in Sec.~\ref{Sec3}, for constructing DBs and for studying their stability. The most widely known method for constructing DBs was developed in \cite{Marin-Aubry}. It based on the Newton-Rafson iterative scheme for finding initial values of all dynamical variables which lead to localized and time-periodic solution of the dynamical equations describing the considered model. However, such approach need, as a rule, very good initial approximation for the breathers profile at $t = 0$ and, even in this case, some problems with convergence of the numeric procedure in the many-dimensional phase space PS can arise. We prefer to construct DBs with the aid of a certain variant of the steepest descent method in the above mentioned space PS (see, for example, \cite{Flach4}). Using this method, we minimize the sum of square deviations between all dynamical variables at the instants $t = 0$ and $t = T$ where $T$ is an arbitrary fixed period of the breathers solution. In some cases, we used the "pair synchronization method" \cite{ChechinBez} in a manual regime for constructing plausible initial conditions. Runge-Kutta methods (rkf45 and dverk78) were applied for integrating differential equations of the considered model from $t = 0$ to $t = T$. Such approach allows us to construct discrete breathers with a high level of accuracy for different scalar models.

We used the same invariant manifolds from Table 1 for constructing DBs for both model 1 and model 2. For example, for the manifold $Q^{(1)}_{5\times5}$, we obtained the following initial conditions which determine DB with $T = 2$ for the model 2 with $\gamma = 1$ (all initial velocities of the dynamical variables are assumed to  be zero):

\begin{equation}\label{eqB500}
    \begin{array}{c}
                    a(0)=2.23751, \ b(0)= -0.64719, \ c(0)=0.30201, \ d(0)=0.13725,\\
                    e(0)=-0.07879, \ f(0)=0.02294.\\
     \end{array}
\end{equation}

On the other hand, using the above discussed method for the model 1 ($\gamma=1$, \ $m=4$), we found the following DB with $T=2$ on the same manifold $Q_{5\times 5}^{(1)}$:

\begin{equation}\label{eqB501}
    \begin{array}{c}
                    a(0)=1.28198, \ b(0)=-0.28941, \ c(0)=0.00366, \ d(0)=0.00180,\\
                    e(0)=-0.00018, \ f(0)\sim 10^{-14}.\\
     \end{array}
\end{equation}

\noindent
(This breather was also obtained with the aid of the Rosenberg nonlinear normal modes technique discussed in Sec.~\ref{Sec3}).

Comparing solutions (\ref{eqB500}) and (\ref{eqB501}), we see that discrete breather in the homogeneous potential model 1 turns out to be much more localized in space than that in the model 2.

Because of this reason, DBs on the invariant manifolds $Q_{5\times 5}^{(1)}$ and  $Q_{5\times 5}^{(2)}$ for the model 2 are essentially different, while DBs on the same manifolds for the model 1 prove to be practically identical. Indeed, the peripheral dynamical variables are very small in this case and the breathers on $Q_{5\times 5}^{(1)}$ and  $Q_{5\times 5}^{(2)}$ manifolds are practically \emph{indistinguishable} from each other up to the used accuracy (note that $Q_{5\times 5}^{(2)}$ transforms into $Q_{5\times 5}^{(1)}$ if $g=f=e$).

\begin{figure}[htb]
\begin{center}
\includegraphics[scale=0.79]{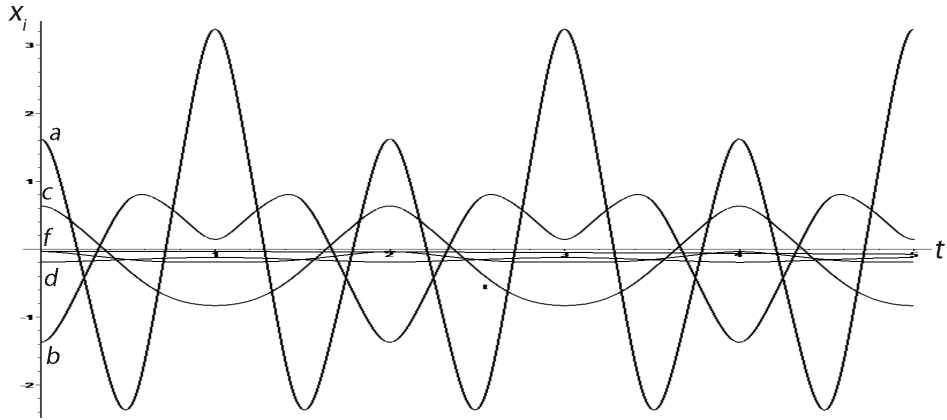}
\caption{Time evolution of DB for the model 1 on the invariant manifold $Q^{(1)}_{5\times 5}$. This breather is not a Rosenberg nonlinear normal mode. }{\label{PPbreather} }
\end{center}
\end{figure}

Surprisingly, using the above described approach, we were able to find some "unusual" DBs both for the models 1 and 2. As an example, in Fig.~\ref{PPbreather}, we present DB with $T=2$ on the manifold $Q_{5\times 5}^{(1)}$ for the model 1 with homogeneous potential of $m=4$ degree. Note that this discrete breather is not a Rosenberg nonlinear normal mode. Indeed, it can be seen in Fig.~\ref{PPbreather} that different particles pass their equilibrium position at \emph{different} instants and this contradicts to the definition of the Rosenberg NNMs [see Eq.~(\ref{eq30b})].  As to our knowledge, all papers devoted to discrete breathers in the models with \emph{homogeneous} potentials deal only with the solutions constructing on the basis of separation of space and time variables.  Such solutions represent Rosenberg nonlinear normal modes. However, one can see in Fig.~\ref{PPbreather} that in the models with homogeneous potentials there can also exist essentially another types of discrete breathers.

The unusual DB presented in Fig.~\ref{PPbreather} is determined by the following initial values of dynamical variables:

\begin{equation*}
    a(0)=1.6208, \ b(0)=-1.3718, \ c(0)=0.6325, \ d(0)=-0.0370, \ f(0)=-0.0354,
\end{equation*}

\begin{figure}[htb]
\begin{center}
\includegraphics[scale=1.15]{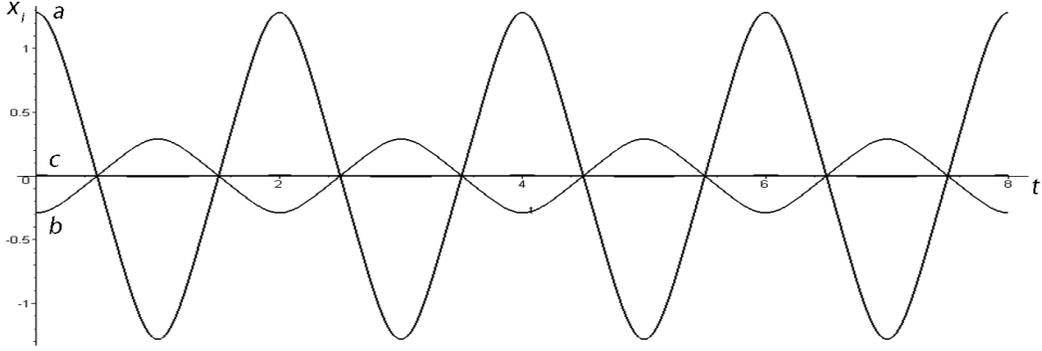}
\caption{Time evolution of DB for the model 1 on the invariant manifold $Q^{(1)}_{5\times 5}$. This breather is a Rosenberg nonlinear normal mode. }{\label{PPG2} }
\end{center}
\end{figure}

\noindent while the breather depicted in Fig.~\ref{PPG2} is generated by Eq.~(\ref{eqB501}).

\noindent
In both cases, initial velocities of all the particles are equal to zero.

Moreover, we can also construct DBs of the above discussed unusual type for the model 2, as well as for other scalar models on square lattice.

\section{General group-theoretical method for simplifying stability analysis of periodic and quasiperiodic regimes }{\label{Sec5}}

\subsection{General discussion}{\label{GD}}

The system (\ref{eqB200}) of variational equations for the model 1 possesses a very specific structure which permits one to decompose it into independent scalar equations by diagonalyzing the matrix $\vec{B}$. Unfortunately, the variational equations for studying stability of breathers  in the model 2 possess more complicated structure which does not allow such decoupling. Indeed, the Jacoby matrix $\vec{J}(t)$ from the variational equations $\ddot{\vec{\delta}}=\vec{J}(t)\vec{\delta}$ for the model 2 can be written as follows
\begin{equation*}
\vec{J}(t)=\vec{D}(t)+\vec{C}.
\end{equation*}
Here $\vec{D}(t)$ is a certain diagonal matrix whose elements are \emph{different} time-periodic functions determined by the breather solution, while $\vec{C}$ is a constant matrix depending on the parameter $\gamma$ which describes coupling between Duffing oscillators.

It is essential, that these two matrices, $\vec{D}(t)$ and $\vec{C}$, don't commute with each other:
\begin{equation*}
\vec{D}(t)\cdot\vec{C}\neq \vec{C}\cdot \vec{D}(t).
\end{equation*}
Therefore, it is impossible to diagonalize them \emph{simultaneously}, i.e. with the aid of one and the same orthogonal transformation.

In turn, this means that one cannot decompose the system of variational equations \linebreak $\ddot{\vec{\delta}}=\vec{J}(t)\vec{\delta}$  into independent scalar equations.

Then, one may ask: "Are there any methods for decomposing the system of variational equations into some independent subsystems of \emph{less} dimension than that of the whole system $\ddot{\vec{\delta}}=\vec{J}(t)\vec{\delta}$"? A general group-theoretical method was developed for solving this problem in our paper \cite{Chechin-Zhukov}. Below, we give an outline of this method adapted to the 2D scalar models.

Let us suppose that the system of nonlinear dynamical equations (\ref{eqv1}) is invariant with respect to a group $G_0$. This means that after the action of induced operators $\hat{g}$, associated with each $g\in G_0$, we obtain a new system of differential equations which is equivalent to the original system (\ref{eqv1}).

It can be seen from the structure of Eqs.~(\ref{eqv1}) and the group $G_0$ (\ref{eq10}) that under the action of any operator $\hat{g}$ ($g\in G_0$) the individual equations are transposed and change their sings exactly in such a way, as the corresponding dynamical variables $q_{ij}$.

On the other hand, as was discussed in Sec.~\ref{seq1}, every dynamical regime in the considered model can be associated with a certain subgroup $G$ of the parent symmetry group $G_0$ \ ($G\subset G_0$). For studying its linear stability, we linearize nonlinear equations (\ref{eqv1}) near this regime and obtain a variational system $\ddot{\vec{\delta}}=\vec{J}(t)\vec{\delta}$, where $\vec{\delta}=\vec{\delta}(t)$ is an infinitesimal vector describing perturbations of all the variables $q_{ij}(t)$, while $\vec{J}(t)$ is the Jacoby matrix of the system (\ref{eqv1}).

Then the following question arises: "What one can say about symmetry group of the linearized (variational) system $\ddot{\vec{\delta}}=\vec{J}(t)\vec{\delta}$?". It was proved in \cite{Chechin-Zhukov} that this system is invariant with respect to the group $G$ describing the symmetry of the given dynamical regime.

Since each operator $\hat{g}$ generates a transposition (and, may be, changing in sign) of individual variables $q_{ij}$, one can associate a certain matrix with every $g\in G$. In this way, we can construct a matrix representation $\Gamma$ of the group $G$. In general, this representation turns out to be reducible and, therefore, it can be decomposed into a number of irreducible representations (irreps) $\Gamma_j$ of the group $G$ with the aid of standard group-theoretical methods (see, for example, \cite{Elliot}):

\begin{equation}\label{eqB100}
    \Gamma=\sum_{j} \phantom{1}\!\!\!^{\oplus}\Gamma_j,
\end{equation}

In \cite{Chechin-Zhukov}, we proved that the Jacoby matrix $J(t)$ commutes with all matrices of the representation $\Gamma$. This fact allows one to apply then the well-known Wigner theorem  \cite{Elliot}. As a consequence of this theorem, the matrix $\vec{J(t)}$ can be represented in the following block-diagonal form
\begin{equation}\label{eqD2}
    \vec{J}(t)=\sum_{j} \phantom{1}\!\!\!^{\oplus} D_j,
\end{equation}
where each block $D_j$ corresponds to the irrep $\Gamma_j$ of the group $G$. Dimension of $D_j$ is equal to $m_j \ \cdot n_j$, where $n_j$ is the dimension of the irrep $\Gamma_j$, while $m_j$ determines how many times $\Gamma_j$ enters into the decomposition (\ref{eqB100}) of the reducible representation $\Gamma$.

Moreover, each block $D_j$ possesses a very specific structure: it consists of subblocks representing matrices proportional to the identity matrix $I_{n_j}$ of dimension $n_j$ which are repeated $m_j$ times along the rows and columns of the block $D_j$.

We can illustrate the structure of a certain block $D_j=D$ characterized by the numbers $n_j=n$, \ $m_j=m$ as follows
\begin{equation}\label{eqD3}
    D=\left(
        \begin{array}{cccc}
          \mu_{11} I_n & \mu_{12} I_{n} & ... & \mu_{1m} I_n  \\
          \mu_{21} I_n & \mu_{22} I_{n} & ... & \mu_{2m} I_n  \\
          ... & ... & ... & ... \\
          \mu_{m1} I_n & \mu_{m2} I_{n} & ... & \mu_{mm} I_n  \\
        \end{array}
      \right),
\end{equation}
where $I_n$ is the $n\times n$ identity matrix.

As was already noted, individual nonlinear equations (\ref{eqv1}), as well as those of linearized system  $\ddot{\vec{\delta}}=\vec{J}(t)\vec{\delta}$  are transformed under the action of operators $\hat{g}$ \ ($g\in G$) as the corresponding variables $q_{ij}$. Therefore, we can construct the same reducible representation $\Gamma$ of the group $G$ using as the basis  the set of equations entering the system $\ddot{\vec{\delta}}=\vec{J}(t)\vec{\delta}$. In turn, this leads to representing this system  in the form of independent subsystems of $m_j \cdot n_j$ equations corresponding to the individual irreps $\Gamma_j$. Moreover, the specific structure (\ref{eqD3}) of the blocks $D_j$ in (\ref{eqD2}) allows one to conclude (see details in \cite{Chechin-Zhukov}) that the subsystem corresponding to the block $D_j$ can be splitted into $n_j$ new independent identical subsystems whose dimension is equal to $m_j$.

Thus, as a result of the decomposition of the variational system $\ddot{\vec{\delta}}=\vec{J}(t)\vec{\delta}$, we can obtain for each irrep $\Gamma_j$, entering  the representation $\Gamma$, \ $n_j$ identical subsystems  whose dimension is equal to $m_j$.

Obviously, such a decomposition may be very effective, if the integer numbers $m_j$ are much smaller than the full dimension of the system $\ddot{\vec{\delta}}=\vec{J}(t)\vec{\delta}$.

It is very important that the above described decomposition does not depend on the character of the considered dynamical regime --- it can be periodic, quasiperiodic or even chaotic!

As was discussed in \cite{Chechin-Zhukov}, for decomposing (splitting) the original system $\ddot{\vec{\delta}}=\vec{J}(t)\vec{\delta}$ into independent subsystems in explicit form, we must find the basis vectors of all $\Gamma_j$ entering $\Gamma$. These vectors can be used as columns of the matrix $S$ which allows one to split the system $\ddot{\vec{\delta}}=\vec{J}(t)\vec{\delta}$.

Using this matrix, we can introduce a new infinitesimal vector $\vec{y}$ instead of the old vector $\vec{\delta}$ with the aid of relation
\begin{equation}\label{eqD10}
    \vec{y}=\vec{S}\vec{\delta}.
\end{equation}
Note that $\vec{S}$ is an orthogonal matrix: \ $\vec{S}^T\cdot \vec{S}=\vec{I}$ ($\vec{S}^T$ is the transpose of the matrix $\vec{S}$). Multiplying both sides of the equation  $\ddot{\vec{\delta}}=\vec{J}(t)\vec{\delta}$ by the matrix $\vec{S}$ from the left, we obtain \ $\vec{S}\ddot{\vec{\delta}}=(\vec{S}\vec{J}(t)\vec{S}^T)\cdot(\vec{S}\vec{\delta})$, or
\begin{equation}\label{eqD11}
    \ddot{\vec{y}}=\tilde{\vec{J}}(t)\cdot \vec{y}.
\end{equation}
Here the matrix $\tilde{\vec{J}}(t)\equiv \vec{S}\cdot \vec{J}(t) \vec{S}^T$ possesses the above discussed block-diagonal structure and, therefore, Eqs.~(\ref{eqD11})
represent a set of \emph{independent} subsystems whose dimensions are determined by $m_j$~--- the number of times that the irrep $\Gamma_j$ enters the reducible representation~$\Gamma$.

\subsection{Splitting schemes}

The explicit decomposition of the linearized system $\ddot{\vec{\delta}}=\vec{J}(t)\vec{\delta}$ represents a cumbersome procedure and, therefore, it is interesting to know beforehand to what extent such decomposition will be useful. For this purpose, we construct the so-called splitting schemes \cite{Chechin-Zhukov} with the aid of the theory of characters of group representations. Let us consider this approach in more detail.

The splitting scheme determines how many independent subsystems one can obtain as a result of the decomposition of the linearized system $\ddot{\vec{\delta}}=\vec{J}(t)\vec{\delta}$ and what dimension possesses each of these subsystems.

To obtain the splitting scheme of variational equations corresponding to a given dynamical regime, we must find how  many times ($m_j$) each irrep $\Gamma_j$ enters the reducible representation $\Gamma$ associated with the symmetry group $G$ of the considered regime.

According to the theory of group representations \cite{Elliot}
\begin{equation}\label{eqE1}
    m_j=\frac{1}{||G||}\sum_{g\in G} \chi_\Gamma (g)\cdot \bar{\chi}_j (g)
\end{equation}
Here $||G||$ is the number of elements of the group $G$, \ $\chi_{\Gamma}(g)$ \ and \ $\chi_{j}(g)$ are traces of the matrices associated with the group element $g\in G$ in the reducible representation $\Gamma$ and in the irreducible representation $\Gamma_j$, respectively [the bar over $\chi_j (g)$ denotes complex conjugation].

As an example, let us consider constructing of the splitting scheme for a dynamical regime on the invariant manifold $Q_{3\times 3}^{(1)}$ corresponding to the point symmetry group $C_{4v}=\{C_4, \ \sigma_y\}$. This group consists of the following eight elements ($||G||=8$):

 \begin{equation}\label{eqBB100}
    \begin{array}{l}
                    g_1(x, \ y)=(x, \ y)~\mbox{--- identity element};\\
                    g_2(x, \ y)=(-y, \ x)~\mbox{~--- rotation by } 90^\circ;\\
                    g_3(x, \ y)=(-x, \ -y)~\mbox{~--- rotation by } 180^\circ;\\
                    g_4(x, \ y)=(y, \ -x)~\mbox{~--- rotation by } 270^\circ;\\
                    g_5(x, \ y)=(-x, \ y)~\mbox{~--- reflection in the mirror plane (1, 0) } [\sigma_x];\\
                    g_6(x, \ y)=(-y, \ -x)~\mbox{~--- reflection in the mirror plane (-1, 1) } [\sigma_{\bar{x}y}];\\
                    g_7(x, \ y)=(x, \ -y)~\mbox{~--- reflection in the mirror plane (0, 1) } [\sigma_y];\\
                    g_8(x, \ y)=(y, \ x)~\mbox{~--- reflection in the mirror plane (1, 1) } [\sigma_{xy}];\\
                  \end{array}
\end{equation}

Here we define every symmetry elements by its action on an arbitrary point ($x, y$) of the two-dimensional plane. All rotations are performed about $Z$ axis perpendicular to this plane and passing through the origin of the coordinate system. Mirror planes are determined by their  normals which are given as two-dimensional vectors at the end of the corresponding lines.
(In square brackets, we give notations of these planes used in the previous sections of the present paper).

Let us illustrate definitions of the symmetry elements from Eqs.~(\ref{eqBB100}) with the example of $g_2$. This element, being the rotation by the angle $90^\circ$, maps the point $(x,y)$ onto the point $(-y, x)$, i.e. $x\rightarrow -y$, \ $y\rightarrow x$ under the action of $g_2$.

Every dynamical regime on the invariant manifold $Q_{3\times 3}^{(1)}$ is described by three variables $a(t)$, $b(t)$, $c(t)$. However, an arbitrary perturbation of this regime is characterized, obviously, by nine independent variables \ $\delta_{ij}$ \ ($i=1..3$, \ $j=1..3$):
\begin{equation}\label{eq350}
    \delta=\left(
             \begin{array}{ccc}
               \delta_{11} &  \delta_{12}  &  \delta_{13}  \\
               \delta_{21}  & \delta_{22} & \delta_{23} \\
               \delta_{31} & \delta_{32} & \delta_{33} \\
             \end{array}
           \right).
\end{equation}
Thus, we must consider the nine-dimensional space of all possible perturbations.

In this space, we choose a \emph{natural} basis $\{\vec{e}_1, \ \vec{e}_2, ... , \vec{e}_9\}$:
\begin{equation*}
    \vec{e}_1=\left(
                \begin{array}{ccc}
                  1 & 0 & 0 \\
                  0 & 0 & 0 \\
                  0 & 0 & 0 \\
                \end{array}
              \right), \ \ \
     \vec{e}_2=\left(
                \begin{array}{ccc}
                  0 & 1 & 0 \\
                  0 & 0 & 0 \\
                  0 & 0 & 0 \\
                \end{array}
              \right), \ \ \ 
        \vec{e}_3=\left(
                \begin{array}{ccc}
                  0 & 0 & 1 \\
                  0 & 0 & 0 \\
                  0 & 0 & 0 \\
                \end{array}
              \right), \ \ \ ..., \ \ \
    \vec{e}_9=\left(
                \begin{array}{ccc}
                  0 & 0 & 0 \\
                  0 & 0 & 0 \\
                  0 & 0 & 1 \\
                \end{array}
              \right).
\end{equation*}
Only one component of each vector $\vec{e}_j$ possesses nonzero value (it is equal to unity).

Basis vectors $\vec{e}_j$ \ ($j=1..9$) transform into each other under the action of any symmetry element \ $g\in G$.
For example, \ $\hat{g}_7 \vec{e}_1=\vec{e}_7$, \  $\hat{g}_7 \vec{e}_2=\vec{e}_8$, \  $\hat{g}_7 \vec{e}_3=\vec{e}_9$, \ ...,  \ $\hat{g}_7 \vec{e}_9=\vec{e}_3$. This transposition of the basis vectors determines a $9\times 9$ matrix $M(g_7)$ of the reducible \emph{permutational} representation $\Gamma$ corresponding to the element $g_7$. In such manner, we can construct the complete representation $\Gamma$ as the ordered set of matrices \ $\{M(g) \ | \ \forall g\in G\}.$

However, for using Eq.~(\ref{eqE1}), we need only the character of the representation $\Gamma$, i.e. the ordered set of traces $\chi(g)$ of all its matrices $M(g)$.

It is easy to see that nonzero contribution to $\chi(g)$ originates only from the basis vectors \ $\vec{e}_j$ which are not changed under the action of the element $g$. For $g_7$ such vectors contain 1 in those sites of the lattice which lie on this mirror plane: \ $\hat{g}_7 \vec{e}_4=\vec{e}_4$, \  $\hat{g}_7 \vec{e}_5=\vec{e}_5$ \ and \ $\hat{g}_7 \vec{e}_6=\vec{e}_6$. Therefore, \ $\chi(g_7)=3$.

Continuing in such a way, we obtain traces $\chi(g)$ for all $g\in G=C_{4v}$ for the reducible representation $\Gamma$ (see the last line in Table 2).

Table 2.

Irreducible representations $\Gamma_j$ \ ($j=1..5$) of the group $G=C_{4v}$ and the character $\chi (\Gamma)$ of its permutational representation $\Gamma$.

\noindent
\begin{tabular}{|c|c|c|c|c|c|c|c|c|}
  \hline
  % after \\: \hline or \cline{col1-col2} \cline{col3-col4} ...
    & $g_1$ & $g_2$ & $g_3$ & $g_4$ & $g_5$ & $g_6$ & $g_7$ & $g_8$ \\\hline
  $\Gamma_1$ & 1 & 1 & 1 & 1 & 1 & 1 & 1 & 1 \\\hline
  $\Gamma_2$ & 1 & 1 & 1 & 1 & -1 & -1 & -1 & -1 \\\hline
  $\Gamma_3$ & 1 & -1 & 1 & -1 & 1 & -1 & 1 & -1 \\\hline
  $\Gamma_4$ & 1 & -1 & 1 & -1 & -1 & 1 & -1 & 1 \\\hline
  $\Gamma_5$ & $\left(\begin{array}{cc} 1 & 0 \\ 0 & 1 \\ \end{array} \right)$
  &
  $\left(\begin{array}{cc} 0 & 1 \\ -1 & 0 \\ \end{array} \right)$
  &
  $\left(\begin{array}{cc} -1 & 0 \\ 0 & -1 \\ \end{array} \right)$
  &
  $\left(\begin{array}{cc} 0 & -1 \\ 1 & 0 \\ \end{array} \right)$
  &
  $\left(\begin{array}{cc} -1 & 0 \\ 0 & 1 \\ \end{array} \right)$
  &
  $\left(\begin{array}{cc} 0 & 1 \\ 1 & 0 \\ \end{array} \right)$
  &
  $\left(\begin{array}{cc} 1 & 0 \\ 0 & -1 \\ \end{array} \right)$
  &
  $\left(\begin{array}{cc} 0 & -1 \\ -1 & 0 \\ \end{array} \right)$
  \\\hline
  $\chi (\Gamma_5)$ & 2 & 0 & -2 & 0 & 0 & 0 & 0 & 0
  \\\hline
  $\chi (\Gamma)$ & 9 & 1 & 1 & 1 & 3 & 3 & 3 & 3
  \\
  \hline
\end{tabular}

On the other hand, the group $G=C_{4v}$ possesses five irreducible representations: four of them are one-dimensional ($\Gamma_1$, \ $\Gamma_2$, \ $\Gamma_3$, \ $\Gamma_4$), while $\Gamma_5$ turns out to be two-dimensional. This information is presented in Table 2 and can be found practically in all textbooks on the group theory.

Using Table 2 and Eq.~(\ref{eqE1}), we can easily obtain:
\begin{equation*}
    m(\Gamma_1)=3, \ \ m(\Gamma_2)=0, \ \ m(\Gamma_3)=1, \ \ m(\Gamma_4)=1, \ \ m(\Gamma_5)=2.
\end{equation*}
Therefore, the reducible representation $\Gamma$ can be decomposed into irreducible representations as follows:
\begin{equation}\label{eqE2}
    \Gamma=3\Gamma_1\oplus \Gamma_3\oplus  \Gamma_4\oplus 2 \Gamma_5.
\end{equation}

According to the above-discussed method for decoupling the linearized system  $\ddot{\vec{\delta}}=\vec{J}(t)\vec{\delta}$  into some independent subsystems, we, therefore, find the following splitting scheme:
\begin{equation}\label{eqE3}
    (3\times 3){\oplus} (1\times 1)\oplus  (1\times 1) \oplus 2 (2\times 2).
\end{equation}

Thus, $9\times 9$ system of variational equations $\ddot{\vec{\delta}}=\vec{J}(t)\vec{\delta}$  can be decomposed into five independent subsystems: 

-- one of them is three-dimensional (it  corresponds to the irrep $\Gamma_1$), 

-- two represent \emph{different} scalar equations, i.e. $1\times 1$ system, corresponding to the irreps $\Gamma_3$ and $\Gamma_4$,  

--
 two subsystems are two-dimensional with the same $2\times 2$ matrix (they correspond to the irrep $\Gamma_5$).

Obviously, the study of stability of the considered dynamical regime becomes considerably easer: instead of analyzing $9\times 9$ system of variational equations one can study stability properties of the four independent subsystems whose dimensions are equal to 1, 2 and 3. Such approach demonstrates its particular effectiveness in the cases where dynamical regimes are \emph{quasiperiodic} and, therefore, we cannot use the Floquet method for studying their stability.

\subsection{Explicit decomposition of the variational equations}

Now we will consider the method for obtaining the \emph{explicit form} of the decoupled subsystems of the linearized system  $\ddot{\vec{\delta}}=\vec{J}(t)\vec{\delta}$.

As was already discussed, to this end, we must find the orthogonal matrix whose columns are basis vectors of all the irreps $\Gamma_j$ entering the reducible representation $\Gamma$. Usually, basis vectors of irreducible representations of a given group are constructed by the projection operator method \cite{Elliot}. However, we prefer to use the "direct" method \cite{chechin} based on the straightforward
application of the group representation definition. This method turns out to be more obvious and more simple for our purposes.

Suppose $\Phi=\{\vec{V}_1, \ \vec{V}_2, \ ..., \ \vec{V}_n\}$ is a "supervector" representing the set of all basis vectors $\vec{V}_j$ of a given (reducible or irreducible) $n$-dimensional representation $\tilde{\Gamma}$ of the group $G$:
\begin{equation*}
    \tilde{\Gamma}=\{M(g)| \forall g\in G\}.
\end{equation*}
This means that (see, for example, \cite{Elliot}):
\begin{equation}\label{eq400}
    \hat{g} {\vec{\Phi}}=M^T(g) {\vec{\Phi}}, \ \ \forall g\in G,
\end{equation}
where $M^T(g)$ is the transpose of the matrix $M(g)$ associated with the element  $g$ of the group $G$.

We want to construct basis vectors of all irreps $\Gamma_j$ \ ($j=1..5$) of the group $G=C_{4v}$ presented in Table 2.
Since $\Gamma_1$, \ $\Gamma_2$, \ $\Gamma_3$ and $\Gamma_4$ are one-dimensional, only one basis vector corresponds to each of them. We denote these vectors by $\vec{\varphi}_1$, \ $\vec{\varphi}_2$, \  $\vec{\varphi}_3$ and  $\vec{\varphi}_4$, respectively.

Note that it is sufficient to take into account only those equations from (\ref{eq400}) which correspond to \emph{generators} of the group $G$: all other elements can be constructed as a different products of these generators and equations (\ref{eq400}), written for such  elements $g\in G$, don't give us any additional information about the basis vectors of a given representation. We choose elements $g_2$ and $g_5$ as generators of the considered group $G=C_{4v}$.

Then, from the definition (\ref{eq400}), we obtain the following equations for the above basis vectors:
\begin{equation}\label{eq450}
    \Gamma_1: \ \hat{g}_2 \vec\varphi_1 = \varphi_1, \ \ \hat{g}_5{\vec{\varphi_1}}=\vec\varphi_1;
\end{equation}

\begin{equation}\label{eq451}
    \Gamma_2: \ \hat{g}_2 \vec\varphi_2 = \varphi_2, \ \  \hat{g}_5{\vec{\varphi_2}}=-\vec\varphi_2;
\end{equation}

\begin{equation}\label{eq452}
    \Gamma_3: \ \hat{g}_2 \vec\varphi_1 =- \varphi_3, \ \  \hat{g}_5{\vec{\varphi_1}}=\vec\varphi_3;
\end{equation}

\begin{equation}\label{eq453}
    \Gamma_4: \ \hat{g}_2 \vec\varphi_4 = -\varphi_4, \ \  \hat{g}_5{\vec{\varphi_4}}=-\vec\varphi_4;
\end{equation}

On the other hand, the irrep $\Gamma_5$ is two-dimensional and, therefore, two basis vectors, $\vec\psi_1$ and $\vec\psi_2$, correspond to it. They satisfy the following equations:
 \begin{equation}\label{eq454}
    \begin{array}{c}
                    \hat{g_2}\vec\psi_1=\vec\psi_2, \ \ \hat{g_2}\vec\psi_2=-\vec\psi_1;\\
                    \hat{g_5}\vec\psi_1=-\vec\psi_1, \ \ \hat{g_5}\vec\psi_2=\vec\psi_2.\\
                  \end{array}
\end{equation}

Solving Eqs. (\ref{eq450})--(\ref{eq454}), we can construct the complete set of basis vectors in nine-dimensional space  of all possible deviations $\delta_{ij}(t)$ [see Eq.~(\ref{eq350})] from the considered dynamical regime on the invariant manifold $Q_{3\times 3}^{(1)}$.

Noting that Eqs.~(\ref{eq450}) coincide with Eqs.~(\ref{eq20})--(\ref{eq211}) up to renaming variables ($q_{ij}\rightarrow\delta_{ij}$), we conclude that the basis vector of the identical irrep $\Gamma_1$ possesses the same form as the invariant manifold $Q_{3\times 3}^{(1)}$ in Eq.~(\ref{eq50a}):
\begin{equation}\label{eq500}
    \vec{\varphi}_1=\left(
                      \begin{array}{ccc}
                        C & B & C \\
                        B & A & B \\
                        C & B & C \\
                      \end{array}
                    \right).
\end{equation}

Thus, $\vec{\varphi}_1$ depends on three arbitrary parameters $A$, $B$, $C$ and represents a three-dimensional subspace of the above-mentioned nine-dimensional space. This means that the irrep $\Gamma_1$ enters the reducible representation $\Gamma$ \emph{three} times.

Taking into account Eq.~(\ref{eq500}), we can rewrite $\vec{\varphi}_1$ as follows:
\begin{equation}\label{eq501}
    \vec{\varphi}_1=A\left(
                       \begin{array}{ccc}
                         0 & 0 & 0 \\
                         0 & 1 & 0 \\
                         0 & 0 & 0 \\
                       \end{array}
                     \right) + B \left(
                       \begin{array}{ccc}
                         0 & 1 & 0 \\
                         1 & 0 & 1 \\
                         0 & 1 & 0 \\
                       \end{array}
                     \right) + C \left(
                       \begin{array}{ccc}
                         1 & 0 & 1 \\
                         0 & 0 & 0 \\
                         1 & 0 & 1 \\
                       \end{array}
                     \right).
\end{equation}
Therefore, one can choose the following three basis vectors $\vec{V}_j$ in the three-dimensional subspace determined by $\vec{\varphi}_1$:
\begin{equation}\label{eq502}
                      \begin{array}{ccc}
                      \vec{V}_1=(0,0,0|0,1,0|0,0,0)^T; \\
                      \vec{V}_2=\frac{1}{2}(0,1,0|1,0,1|0,1,0)^T;  \\
                      \vec{V}_3=\frac{1}{2}(1,0,1|0,0,0|1,0,1)^T. \\
                      \end{array}
\end{equation}
These nine-dimensional vectors form the first three columns of the matrix $S$ which leads to the decomposition of the linearized system $\ddot{\vec{\delta}}=J(t)\cdot \vec{\delta}$ into some independent subsystems. (Here we use the transposition symbol $T$ to present column vectors as row vectors).

From Eqs.~(\ref{eq451}), we find $\vec{\varphi}_2=0$, i.e. the irrep $\Gamma_2$ does not enter the reducible representation $\Gamma$.

The solution
\begin{equation}\label{eq503}
    \vec{\varphi}_3=\left(
                      \begin{array}{ccc}
                        0 & -B & 0 \\
                        B & 0 & B \\
                        0 & -B & 0 \\
                      \end{array}
                    \right)
\end{equation}
of Eqs.~(\ref{eq452}) means that the irrep $\Gamma_3$ enters the representation $\Gamma$ only one time and, therefore, the fourth column of the matrix $S$ represents the vector
\begin{equation*}
    \vec{V}_4=\frac{1}{2}(0,-1,0|1,0,1|0,-1,0)^T.
\end{equation*}

From the solution of Eq.~(\ref{eq453})
\begin{equation}\label{eq503}
    \vec{\varphi}_4=\left(
                      \begin{array}{ccc}
                        -A & 0 & A \\
                        0 & 0 & 0 \\
                        A & 0 & -A \\
                      \end{array}
                    \right)
\end{equation}
we obtain the fifth column of the matrix $S$:
\begin{equation*}
    \vec{V}_5=\frac{1}{2}(-1,0,1|0,0,0|1,0,-1)^T.
\end{equation*}

Finally, for the irrep $\Gamma_5$, we find from Eq.~(\ref{eq454}):
\begin{equation}\label{eq505}
    \vec{\psi}_1=\left(
                      \begin{array}{ccc}
                        A & B & A \\
                        0 & 0 & 0 \\
                        -A & -B & -A \\
                      \end{array}
                    \right), \ \ \ \
    \vec{\psi}_2=\left(
                      \begin{array}{ccc}
                        A & 0 & -A \\
                        B & 0 & -B \\
                        A & 0 & -A \\
                      \end{array}
                    \right).
\end{equation}
These basis vectors depend on two parameters ($A, \ B$) and, this confirms that the irrep $\Gamma_5$ enters the representation $\Gamma$ two times [see Eq.~(\ref{eqE2})]. Supposing successively \ $A=1$, \ $B=0$ \ and  \ $A=0$, \ $B=1$, we obtain after normalizing the last four columns of the matrix $S$:
\begin{equation*}
                      \begin{array}{ccc}
                      \vec{V}_6=\frac{1}{2}(1,0,1|0,0,0|-1,0,-1)^T; \\
                      \vec{V}_7=\frac{1}{\sqrt{2}}(0,1,0|0,0,0|0,-1,0)^T;  \\
                      \vec{V}_8=\frac{1}{2}(1,0,-1|0,0,0|1,0,-1)^T; \\
                      \vec{V}_9=\frac{1}{\sqrt{2}}(0,0,0|1,0,-1|0,0,0)^T. \\
                      \end{array}
\end{equation*}

All the vectors $\vec{V}_j \ (j=1..9)$ are orthogonal and {\it normalized}. Note that this is the necessary condition for the matrix $\vec{S}$ to be orthogonal.

As a result of all above-discussed steps, we confirm the correctness of the decomposition (\ref{eqE2}) of the reducible representation $\Gamma$ into the  irreps of the group $G=C_{4v}$ and construct the following matrix $S$ which allow us to split the linearized system $\ddot{\vec{\delta}}=J(t)\cdot \vec{\delta}$ into independent subsystems:

\begin{equation}\label{eq600}
    S=\left(
        \begin{array}{ccccccccc}
          \frac12 & 0 & 0 & 0 & -\frac12 & \frac12 & \frac12 & 0 & 0 \\
          0 & \frac12 & 0 & -\frac12 & 0 & 0 & 0 & \frac{\sqrt{2}}{2}  & 0 \\
          \frac12 & 0 & 0 & 0 & \frac12 & \frac12 & -\frac12 & 0 & 0 \\
          0 & \frac12 & 0 & \frac12 & 0 & 0 & 0 & 0 & \frac{\sqrt{2}}{2} \\
          0 & 0 & 1 & 0 & 0 & 0 & 0 & 0 & 0 \\
          0 & \frac12 & 0 & \frac12 & 0 & 0 & 0 & 0 & -\frac{\sqrt{2}}{2} \\
          \frac12 & 0 & 0 & 0 & \frac12 & -\frac12 & \frac12 & 0 & 0 \\
          0 & \frac12 & 0 & -\frac12 & 0 & 0 & 0 & -\frac{\sqrt{2}}{2} & 0 \\
          \frac12 & 0 & 0 & 0 & -\frac12 & -\frac12 & -\frac12 & 0 & 0 \\
        \end{array}
      \right).
\end{equation}

As was described in Sec.~\ref{GD} [see, Eqs.~(\ref{eqD10})--(\ref{eqD11})], introducing the new nine-dimensional infinitesimal vector \ $\vec{y}=S\vec{\delta}$ with matrix $S$ determined by Eq.~(\ref{eq600}), we obtain from the old linearized system $\ddot{\vec{\delta}}=J(t)\cdot \vec{\delta}$  the new one
\begin{equation}\label{eq601}
    \ddot{\vec{y}}=\tilde{J}(t)\vec{y}, \ \ \ \ \tilde{J}(t)=S J(t) S^T,
\end{equation}
which turns out to be decomposed into some independent subsystems.

\textit{Example 1}

For the model (\ref{eq21}) of linearly coupled Duffing oscillators, we obtain the following subsystems corresponding to the individual irreps of the group $G=C_{4v}$.

The irrep $\Gamma_1$ generates the following $3\times 3$ subsystem:
\begin{equation}\label{eq610}
    \left\{         \begin{array}{lll}
                       \ddot{y}_1+[3c^2(t)+(1+2\alpha)]y_1-2\alpha y_2=0,\\
                       \ddot{y}_2-2\alpha y_1+[3b^2(t)+(1+3\alpha)]y_2-2\alpha y_3=0, \\
                       \ddot{y}_3-2\alpha y_2+[3a^2(t)+(1+4\alpha)]y_3=0.\\
                      \end{array}
                    \right.
\end{equation}

All three time-periodic functions, $a(t)$, \ $b(t)$ and $c(t)$, describing a dynamical regime on the invariant manifold $Q_{3\times 3}^{(1)}=\left(
                      \begin{array}{ccc}
                        c & b & c \\
                        b & a & b \\
                        c & b & c \\
                      \end{array}
                    \right)$, enter Eqs.~(\ref{eq610}). Let us emphasize once more that this regime can be periodic, as well as \emph{quasiperiodic}.

Two \emph{identical} $2\times 2$ subsystems
\begin{equation}\label{eq611}
    \left\{         \begin{array}{lll}
                       \ddot{y}_1+[3c^2(t)+(1+4\alpha)]y_1-\sqrt{2}\alpha y_2=0,\\
                       \ddot{y}_2-\sqrt{2}\alpha y_1+[3b^2(t)+(1+5\alpha)]y_2=0\\
                  \end{array}
                    \right.
\end{equation}
correspond to the irrep $\Gamma_5$, while different scalar equations correspond to the irreps $\Gamma_3$ and $\Gamma_4$, namely,
\begin{equation}\label{eq612}
    \Gamma_3: \ \ \ \ddot{y}+[3b^2(t)+(1+3\alpha)]y=0,
\end{equation}

\begin{equation}\label{eq613}
    \Gamma_4: \ \ \ \ddot{y}+[3c^2(t)+(1+6\alpha)]y=0.
\end{equation}

For studying stability of discrete breathers in the model 2, we can apply the standard Floquet method. In the case of the invariant manifold $Q_{3\times 3}^{(1)}$, one has to deal with the 18-dimensional phase space corresponding to a given breather. On the other hand, after splitting the linearized system $\ddot{\vec{\delta}}=J(t)\cdot \vec{\delta}$ into independent subsystems (\ref{eq610})--(\ref{eq613}), we may analyze discrete breather stability applying the Floquet method successively for these subsystems of much smaller dimensions.

For DB with $T=2$ determined on the invariant manifold $Q_{3\times 3}^{(1)}$ in the case of the model 2 with $\gamma=1$ by the initial conditions, \ $a(0)=2.35593$, \ $b(0)=-0.45156$, \ $c(0)=0.13152$ we have obtained the following Floquet exponents for $3\times 3$ subsystem (\ref{eq610}):
\begin{equation}\label{eq700}
        \left\{         \begin{array}{lll}
                       -0.74678\pm0.66507\cdot i,\\
                       0.26213\pm0.99966\cdot i,\\
                       1\pm0\cdot i.\\
                       \end{array}
                    \right.
\end{equation}

For $2\times 2$ subsystems (\ref{eq611}):
\begin{equation}\label{eq701}
        \left\{         \begin{array}{lll}
                       -0.60978\pm0.79257\cdot i,\\
                       0.61449\pm0.78892\cdot i.\\
                       \end{array}
                    \right.
\end{equation}

For Eq.~(\ref{eq612}):
\begin{equation}\label{eq702}
      -0.53267\pm0.84632\cdot i.
\end{equation}

For Eq.~(\ref{eq613}):
\begin{equation}\label{eq703}
      -0.55546\pm0.83154\cdot i.
\end{equation}

All these Floquet exponents lie on the unit circle in the complex plane and, therefore, the given DB proves to be stable.

To check the above-discussed splitting procedure, we calculated the Floquet exponents for the whole 18-dimensional phase space associated with the original linearized system $\ddot{\vec{\delta}}=J(t)\vec{\delta}$ and are convinced that they exactly coincide with those from Eqs.~(\ref{eq700})--(\ref{eq703}).

Actually Eqs.~(\ref{eq700})--(\ref{eq703}) allow us to study stability of a given DB with respect to different symmetry-determined collective degrees of freedom.  In some a sense, such information can throw light upon the cause of the breather stability loss.

\textit{Example 2}

Finally, let us decompose the linearized system $\ddot{\vec{\delta}}=J(t)\vec{\delta}$  for the discrete breather on $Q_{3\times 3}^{(1)}$ in the case of the model 1 whose stability we were able to study without applying the Floquet method (see Sec.~\ref{Sec3}). Using the  \emph{same} matrix $S$ from Eq.~(\ref{eq600}) we obtain the following decomposition of the linearized system:
\begin{equation}\label{eq620}
     \Gamma_1: \ \ \left\{   \begin{array}{lll}
                   \ddot{y}_1+3[\gamma c^2(t)+2\nu(t)]y_1-6\nu(t)y_2=0,  \\
                   \ddot{y}_2-6\nu(t) y_1+3[\gamma b^2(t) + \mu(t)+2\nu(t)]y_2-6\mu(t) y_3=0,  \\
                   \ddot{y}_3-6\mu(t)y_2+3[\gamma a^2(t)+4\mu(t)]y_3=0, \\
                   \end{array}\right.
\end{equation}

\begin{equation}\label{eq621}
     \Gamma_5: \ \ \left\{  \begin{array}{lll}
                   \ddot{y}_1+3[\gamma c^2(t)+2\nu(t)]y_1-3\sqrt{2}\nu(t)y_2=0,  \\
                   \ddot{y}_2-3\sqrt{2}\nu(t) y_1+3[\gamma b^2(t) + \mu(t)+2\nu(t)]y_2=0,  \\
                   \end{array}\right.
\end{equation}

\noindent
(two identical systems of this form),

\begin{equation}\label{eq622}
     \Gamma_3: \ \  \begin{array}{ccc}
                   \ddot{y}+3[\gamma b^2(t)+\mu(t)+2\nu(t)]y=0,  \\
                   \end{array}
\end{equation}

\begin{equation}\label{eq623}
     \Gamma_4: \ \  \begin{array}{ccc}
                   \ddot{y}+3[\gamma c^2(t)+2\nu(t)]y=0.  \\
                   \end{array}
\end{equation}
Here $\mu(t)=[a(t)-b(t)]^2$, \ $\nu(t)=[b(t)-c(t)]^2$, where $a(t)$, \ $b(t)$, and $c(t)$ are three breather components on the invariant manifold $Q_{3\times 3}^{(1)}$.

As a consequence of this decomposition, we can associate different stability indicators $\Lambda_j$ \ ($j=0..8$) \ (see Sec.~\ref{Sec3}) for DB  ($\gamma=1$) with the irreducible representations \ $\Gamma_1$, $\Gamma_3$, $\Gamma_4$, $\Gamma_5$ of the group $G=C_{4v}$:

\begin{equation}\label{eq625}
      \begin{array}{lll}
              \Gamma_1: \ \ \Lambda_0= 3, \ \ \Lambda_1= 0.12407, \ \ \Lambda_2= 0.02340  \\
              \Gamma_3: \ \ \Lambda_3=  0.59433    \\
              \Gamma_4: \ \ \Lambda_4=  0.03707     \\
              \Gamma_5: \ \ \Lambda_5= \Lambda_6= 0.59556, \ \  \Lambda_7=\Lambda_8=0.03584.\\
                   \end{array}
\end{equation}

Note that some of these $\Lambda_j$ turn out to be degenerate, exactly or approximately. We will consider symmetry-related causes of this phenomenon elsewhere.

In connection with the above group-theoretical method for splitting the linearized system $\ddot{\vec{\delta}}=J(t)\vec{\delta}$ into independent subsystems, one may ask: "How this method works in the case of large lattice fragment?" Indeed, the worse the localization of the breather, the larger fragments of the corresponding invariant manifolds one must consider. Unfortunately, our decomposition method becomes less effective for this case because dimensions of the independent subsystems can be rather large. For example, we obtain the following splitting scheme for studying stability of DB with symmetry group $G=C_{4v}$ on the invariant manifold $Q_{5\times 5}^{(1)}$:
\begin{equation}\label{eq800}
    1(6\times 6) \oplus 1(1\times 1) \oplus 1(3\times 3) \oplus1(3\times 3)\oplus2(6\times 6).
\end{equation}
These subsystems are associated with the irreps \ $\Gamma_1$, $\Gamma_2$, $\Gamma_3$, $\Gamma_4$ and $\Gamma_5$, respectively.

Comparing (\ref{eq800}) and (\ref{eqE3}), we see that the decomposition of the linearized system on the manifold $Q_{3\times 3}^{(1)}$ is more efficient than that on $Q_{5\times 5}^{(1)}$.

However, let us emphasize again that all discussed decompositions are true not only for periodic, but for any quasiperiodic regimes, as well. In practice, the latter can be very important.

\section{Conclusion}

In the present paper, we study stationary discrete breathers in two nonlinear scalar models on the plane square lattice. However, the developed group-theoretical methods for constructing breather solutions and studying their stability can be naturally extended to the case of  time-periodic and quasiperiodic nonlinear dynamical objects of various physical nature on different 2D and 3D lattices. These methods allow one to simplify, sometimes considerably, studying discrete breathers and quasibreathers in space-periodic structures.

\section*{Acknowledgments}

The authors are grateful to V.~P.~Sakhnenko and S.~Flach for useful discussions. G.~S.~Bezuglova is very thankful to the Russian foundation "Dynasty" for the financial support.

\end{document}